\documentclass[11pt]{article}
\usepackage[utf8]{inputenc}
\usepackage[english]{babel}
\usepackage{amsmath,amsfonts,amsthm,graphicx,float}
\usepackage{natbib}
\DeclareMathAlphabet{\mathpzc}{OT1}{pzc}{m}{it}

\newtheorem{rem}{Remark}[section]

\usepackage{algorithm}

\usepackage[noend]{algpseudocode}

\def\bm{{\mathbf m}}

\def\b1{{\mathbf 1}}

\usepackage[font=small]{caption}

\newtheorem{theorem}{Theorem}

\usepackage{titlesec}

\titleformat*{\section}{\normalfont\fontsize{14}{17}\bfseries}
\titleformat*{\subsection}{\normalfont\fontsize{12}{15}\bfseries}

\usepackage{bm}
\usepackage{enumerate}
\providecommand{\abs}[1]{\lvert#1\rvert}
\providecommand{\norm}[1]{\lVert#1\rVert}

\usepackage{soul}
\usepackage[usenames, dvipsnames]{color}

\date{}

\begin{document}

\title{\LARGE Nonparametric estimation of circular trend surfaces with application to wave directions}\normalsize
\author{Andrea Meil\'an-Vila \\
Carlos III University of Madrid\thanks{%
Department of Statistics, Carlos III University of Madrid, Av. de la Universidad 30, Legan\'es, 28911, Spain}
\and
Rosa M. Crujeiras \\
%EndAName
Universidade de Santiago de Compostela\thanks{Department of Statistics, Mathematical Analysis and Optimization, Faculty of Mathematics, Universidade de Santiago de Compostela, R\'ua Lope G\'omez de Marzoa s/n,
	15782, Santiago de Compostela, Spain}
\and %
Mario Francisco-Fern\'andez\\
%EndAName
Universidade da Coru\~{n}a\thanks{%
Research group MODES, CITIC, Department of Mathematics, Faculty of Computer Science, Universidade da Coru\~na, Campus de Elvi\~na s/n, 15071,
A Coru\~na, Spain}
}
\maketitle

%.........................................%

\begin{abstract}
In oceanography, modeling wave fields requires the use of statistical tools capable of handling the circular nature of the {data measurements}. An important issue in ocean wave analysis is the study of height and direction waves, being direction values recorded as angles or, equivalently, as points on a unit circle. Hence, reconstruction of a wave direction field on the sea surface can be approached by the use of a linear-circular regression model, viewing wave directions as a realization of a circular spatial process whose trend should be estimated. In this paper, we consider a spatial regression model with a circular response and several real-valued predictors. Nonparametric estimators of the circular trend surface are proposed, accounting for the (unknown) spatial correlation. Some asymptotic results about these estimators as well as some guidelines for their practical implementation are also given. The performance of the proposed estimators is investigated in a simulation study. An application to wave directions in the Adriatic Sea is provided for illustration.

\end{abstract}	
\textit{Keywords:} { Angular risk, Circular data, Local polynomial regression, Spatial correlation, Wave orientation}

%.........................................%

\section{Introduction}
\label{intro}
In many scientific fields, such as oceanography, meteorology or biology, data  are angular measurements (points on the circumference of  the unit circle), exhibiting in some cases a spatial dependence structure which should be accounted for in any modeling approach. For instance,  \cite{casson1998extreme} provided a spatial analysis about the direction of maximum wind speed at locations on the Gulf and Atlantic coasts of the United States. On a series of simulated hurricane wind speeds, the authors aim to model {the stochastic behavior of the extreme wind speeds jointly with their} associated directions. In other scenarios, circular measurements are also accompanied by observations of  real-valued random variables, as in \cite{garcia2014test}, who analyzed the relation between orientation and size of wildfires in Portugal; or \cite{mastrantonio2018distributions}, who proposed a Markov model for multivariate circular-linear data  to forecast the wind speed and direction in the city of Taranto (Italy). Alternative approaches using copulas have been also considered in similar contexts. For instance, \cite{carnicero2013non}, explored the relation between wind direction and rainfall amount in the North of Spain, as well as the dependence between the wind directions in two nearby buoys at the Atlantic ocean.

\begin{figure}
	\centerline{
	\includegraphics[width=8.2cm]{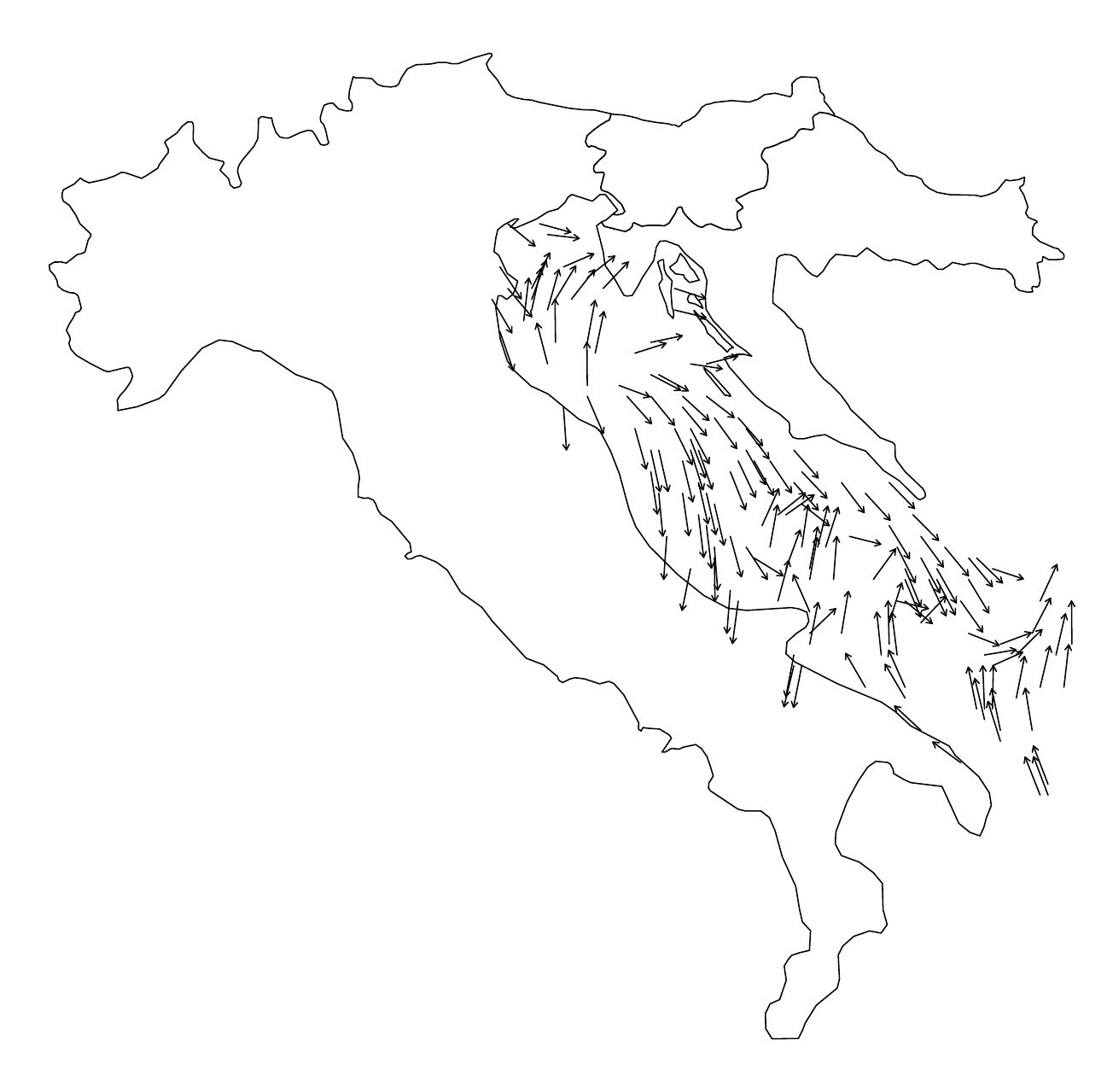}}
	\caption{Random sample of 150 wave directions in the Adriatic Sea area on April 2, 2010 at 6am during a calm period. }			
	\label{figure:realdata}
\end{figure}

In certain situations, the circular data sample is georeferenced, and the goal is to reconstruct the circular trend from a realization of a circular spatial process. This is the case in our motivating example, corresponding to an application in oceanography. Wave directions are recorded in 1494 grid points on the Adriatic Sea area from a calm period transitioning to a storm period at different times. Fig. \ref{figure:realdata} shows a random sample of 150 observations during a calm period. As intuition suggests, these data seem to exhibit a spatial pattern. Considering other periods and moments, this dataset has been deeply studied by several authors using parametric methods. For example,  \cite{jona2012spatial} analyzed outgoing wave directions from a storm period, formulating the wrapped Gaussian spatial process, as a spatial process for circular data.  \cite{mastrantonio2016wrapped}  introduced the wrapped skew Gaussian process as an alternative to the wrapped Gaussian process which allows for asymmetric marginal distributions. This circular process was {also} used {for analyzing} wave directions.  Wave directions from a calm period transitioning to a storm period were also modeled by \cite{wang2014modeling}. They developed the projected Gaussian spatial process, induced from a linear bivariate Gaussian spatial process. Motivated by the same real dataset,  \cite{lagona2015hidden} introduced a hidden Markov model accounting for the  correlation  of spatio-temporal linear-circular data, providing an approach to identify regimes of marine currents.

An alternative to the previous approaches {for modeling} circular data at different spatial locations consists in the estimation of a circular spatial trend using smoothing methods. These techniques relax parametric assumptions of the generating process and, consequently, provide a more flexible way to explore and model the data. This work aims to provide, using smoothing techniques, a circular trend surface estimation procedure. For this purpose, we consider a regression model with a circular response and  an  $\mathbb{R}^d$-valued covariate, assuming that the errors exhibit a spatial correlation structure. For a single real-valued covariate, \cite{di2013non} introduced a nonparametric estimator of the regression function when the  errors  are independent and also when the data come from mixing processes.  The same approach has been also applied, with due modifications, in the context of time series by \cite{di2012non}. The authors considered smoothing and prediction in  the time domain for circular time-series data. Recently,  \cite{meilan2019nonparametric} proposed and studied nonparametric estimators of a circular regression function considering independent data and several real-valued covariates. In the present work, local polynomial-type regression estimators considering a mo\-del with  a circular response and an $\mathbb{R}^d$-valued covariate, in the presence of (unknown) spatial correlation, are introduced and analyzed.

As in any smoothing approach, a crucial step in our proposal  is the selection of an appropriate bandwidth or smoothing parameter {(a symmetric $d\times d$ matrix, for an $\mathbb R^d$-valued covariate in our setting).
This bandwidth matrix controls} the shape and the size
of the local neighborhood used for estimating the spatial trend, which directly impacts the smoothness of the estimator. In the Euclidean setting,  considering a random design and independent data, \cite{ruppert1994multivariate} derived
the asymptotic mean squared error (AMSE) for the multivariate local linear estimator. On the other hand, \cite{opsomer2001nonparametric} provided the corresponding results for  the bidimensional ($d=2$) case when the errors are
correlated, while \cite{liu2001kernel} generalized those results for an arbitrary dimension $d$.  These error expressions allow to derive optimal bandwidths in the corresponding contexts.	 Following similar arguments,   some guidelines to select locally optimal bandwidth matrices are given in this paper. Cross-validation (CV) bandwidth selection methods are also introduced and analyzed in practice. Different simulation scenarios are designed, considering circular spatial errors generated from wrapped and from projected Gaussian spatial processes.

This paper is organized as follows. Section \ref{sec:model}  introduces the linear-circular regression
model, with some highlights on the error process structure.  Nonparametric estimators of the circular regression function are also proposed in this section. Section \ref{sec:est} contains some results about the asymptotic behavior of these estimators. Additionally,  some proposals for bandwidth selection are introduced. A simulation study for assessing
the performance of the estimators and the bandwidth selectors is provided in Section \ref{sec:sim}. These simulations are carried out considering that the errors are drawn from  wrapped (Section \ref{sec:wra}) and from projected (Section \ref{sec:pro}) Gaussian spatial processes. In Section \ref{sec:example}, the application of the proposed approaches to estimate the wave direction trend surface in the {Adriatic Sea} is presented. Finally,  Section \ref{sec:con} contains some conclusions, limitations of the proposal, and {comments on} further research.

%--------------------------------------------------------%
\section{Regression models and estimators}\label{sec:model}
{This section presents the regression models considered in this work and the proposed circular regression estimators. The linear-circular regression model for spatially correlated data is introduced in Section \ref{sec:model_spatial}. {For this model}, nonparametric
	estimators of the circular regression function, based on considering two regression
	models for the sine and cosine components of the response variable, are proposed in Section \ref{sec:estimator}. Specifically, Nadaraya--Watson (NW) and local linear (LL) type nonparametric estimators of the regression function are considered.}

{In what follows, $\bm{\nabla} g(\bm{x})$ and $\bm{\mathcal{H}}_{g}(\bm{x})$ will denote the vector of first-order partial derivatives and  the Hessian matrix of a sufficiently smooth function $g$ at $\bm{x}$, res\-pectively. For a vector $\bm{u}= (u_1,\dots,u_d)^\top$ and an integrable function $g$, the 
	integral $\int\dots\int g(\bm{u}){\textrm{d}}{u}_1\dots{\textrm{d}}{u}_d$ will  be simply denoted as $\int g(\bm{u}){\textrm{d}}\bm{u}$.   Finally, for any
	matrix $\bm{A}$, {$\bm{A}^\top$,} $\abs{\bm{A}}$, ${\rm tr}(\bm{A})$, $\lambda_{{\max}}(\bm{A})$ and $\lambda_{{\min}}(\bm{A})$  denote its {transpose,} determinant,
	trace, maximum eigenvalue and minimum eigenvalue, respectively.}

\subsection{{A regression model with circular response}}\label{sec:model_spatial} Let $\{({\bm{X}}_i,\Theta_i)\}_{i=1}^n$ be a random sample from the $(d+1)$-valued random vector (${\bm{X}}, \Theta$), where $\Theta$ denotes a circular response,  taking values on $\mathbb{T}=[0,2\pi)$, which depends on a $d$-dimensional random variable ${\bm{X}}$, with density $f$ and taking values in  $D\subseteq\mathbb{R}^d$, through the following linear-circular regression model:
\begin{equation}\label{model}
\Theta_i=[m({\bm{X}}_i)+{\varepsilon}_i]({\rm \texttt{mod}}\,2\pi), \quad i=1,\dots,n,
\end{equation} 
where $m$ is a smooth trend or regression function, \texttt{mod} stands for the modulo operation, and $\varepsilon$ denotes a spatially correlated error process {with zero mean direction. Taking into account the definition of the mean of a circular random variable, this condition is equivalent to ${\mathbb{E}}[\sin (\varepsilon)\mid\bm{X}=\bm{x}]=0$. Additionally, we assume that} 
%		\begin{enumerate}[{(C}1)]
		 \begin{eqnarray}
		 {\mathbb{C}\rm ov}[\sin(\varepsilon_i),\sin(\varepsilon_j)\mid{\bm{X}}_i, {\bm{X}}_j]&=&\sigma^2_1{\rho_{1,n}}({\bm{X}}_i-{\bm{X}}_j), \label{cor1}\\
		 {\mathbb{C}\rm ov}[\cos(\varepsilon_i),\cos(\varepsilon_j)\mid{\bm{X}}_i, {\bm{X}}_j]&=&\sigma^2_2{\rho_{2,n}}({\bm{X}}_i-{\bm{X}}_j), \label{cor2}\\
		 {\mathbb{C}\rm ov}[\sin(\varepsilon_i),\cos(\varepsilon_j)\mid{\bm{X}}_i, {\bm{X}}_j]&=&\sigma_{12}{\rho_{3,n}}({\bm{X}}_i-{\bm{X}}_j), \label{cor3}
		 \end{eqnarray}
%		\end{enumerate}
		with    $\sigma^2_k<\infty$, for $k=1,2$, and $\sigma_{12}<\infty$. The continuous stationary correlation functions ${\rho_{k,n}}$ satisfy ${\rho_{k,n}}(\bm{0})=1$, ${\rho_{k,n}}({\bm{x}})={\rho_{k,n}}(-{\bm{x}})$, and $\abs{{\rho_{k,n}}({\bm{x}})}\le1$, for   ${\bm{x}}\in D$, and $k=1,2,3$.  The subscript $n$ in ${\rho_{k,n}}$ indicates that the correlation functions vary with $n$ (specifically, the correlation functions {shrink} as $n$ goes to infinity, as described below). 
		Note also that the subscript $k$ does not correspond to an integer sequence and it just indicates if the correlation corresponds to the sine process ($k=1$), the cosine process ($k=2$) or if it is the cross-correlation between them ($k=3$).
		
		\subsection{{Nonparametric regression estimators}}\label{sec:estimator}
The circular regression function $m$  can be defined as the minimizer of  the usual angular risk ${\mathbb{E}}\{1-\cos[\Theta-m({\bm{x}})]|\bm{X}=\bm{x}\}$. The  solution of this optimization problem is given by{:}
 \begin{equation}
m(\bm{x})=\mbox{arctan2}[m_1(\bm{x}),m_2(\bm{x})],
\label{m}
\end{equation} 
where  {$m_1({\bm{x}})={\mathbb{E}}[\sin(\Theta)\mid{\bm{X}}={\bm{x}}]$, $m_2({\bm{x}})={\mathbb{E}}[\cos(\Theta)\mid{\bm{X}}={\bm{x}}]$, and the function ${\mbox{arctan2}}[y,x]$ returns the angle between the
			$x$-axis and the vector from the origin to $(x, y)$.
			With this formulation, $m_1$ and $m_2$ can be regarded as the regression functions of two regression models  having $\sin(\Theta)$ and $\cos(\Theta)$ as their responses, respectively.
Specifically, we assume the models:
\begin{eqnarray}
		\sin(\Theta_i)&=&m_1({\bm{X}}_i)+\xi_i\label{model1}\\
		\cos(\Theta_i)&=&m_2({\bm{X}}_i)+\zeta_i,\label{model2}\end{eqnarray}
where the $\xi_i$ and the $\zeta_i$  are error terms, absolutely bounded by 1,  satisfying ${\mathbb{E}}(\xi\mid{\bm{X}}={\bm{x}})={\mathbb{E}}(\zeta \mid{\bm{X}}={\bm{x}})=0$. Additionally, for every ${\bm{x}}\in D$, set $s_1^2({\bm{x}})={\mathbb{V}\rm ar}(\xi\mid{\bm{X}}={\bm{x}})$, $s_2^2({\bm{x}})={\mathbb{V}\rm ar}(\zeta\mid{\bm{X}}={\bm{x}})$, $c({\bm{x}})={\mathbb{E}}(\xi\zeta\mid{\bm{X}}={\bm{x}})$, and taking into account that the errors in model (\ref{model}) are spatially correlated, we use the notation  ${\mathbb{C}\rm ov}(\xi_i,\xi_j\mid{\bm{X}}_i, {\bm{X}}_j)={C}_{n,1}({\bm{X}}_i,{\bm{X}}_j)$, ${\mathbb{C}\rm ov}(\zeta_i,\zeta_j\mid{\bm{X}}_i, {\bm{X}}_j)={C}_{n,2}({\bm{X}}_i,{\bm{X}}_j)$ and ${\mathbb{C}\rm ov}(\xi_i,\zeta_j\mid{\bm{X}}_i, {\bm{X}}_j)={C}_{n,3}({\bm{X}}_i,{\bm{X}}_j)$,    for $i,j=1,\dots,n$, and $i\neq j$.
		}

A whole class of kernel-type estimators for $m(\bm{x})$ in (\ref{m})  can be defined replacing in its expression the unknown functions $m_1(\bm{x})$ and $m_2(\bm{x})$ by suitable local polynomial estimators as follows: 
\begin{equation}\label{est}
			\hat{m}_{\bm{H}}(\bm{x};p)=\mbox{arctan2}[\hat{m}_{1, \bm{H}}(\bm{x};p),\hat{m}_{2, \bm{H}}(\bm{x};p)],
\end{equation}
where $\hat m_{1, {\bm{H}}}({\bm{x}};p)$ and $\hat m_{2, {\bm{H}}}({\bm{x}};p)$ denote  the $p$th order local polynomial estimators {(with bandwidth matrix $\bm H$)} of $ m_1({\bm x})$ and $m_2({\bm x})$,  respectively \citep{ruppert1994multivariate,liu2001kernel}. 

Considering $p=0$, the NW estimators of
	the regression functions $m_j$, $j=1,2$, at $\bm{x}\in D$, are respectively defined  as:
\begin{equation*}
\label{estNW}\hat m_{j, \bm{H}}(\bm{x};0)=\left\{\begin{array}{lc}\dfrac{\sum_{i=1}^n K_{\bm{H}}(\bm{X}_i-\bm{x})\sin(\Theta_i)}{\sum_{i=1}^n K_{\bm{H}}(\bm{X}_i-\bm{x})}&\text{if $j=1$},\\\\ \dfrac{\sum_{i=1}^n K_{\bm{H}}(\bm{X}_i-\bm{x})\cos(\Theta_i)}{\sum_{i=1}^n K_{\bm{H}}(\bm{X}_i-\bm{x})}&\text{if $j=2$},\end{array}\right.
\end{equation*}
	where, for $\bm{u}\in\mathbb{R}^d$, $K_{\bm{H}}(\bm{u})=\abs{{\bm{H}}}^{-1}K({\bm{H}}^{-1}\bm{u})$ is the rescaled version of a $d$-variate kernel function $K$, and $\bm{H}$ is a $d\times d$ bandwidth matrix.

	 On the other hand, considering $p=1$, the LL estimators for the regression functions $m_j$, $j=1,2$,  at a given point $\bm{x}\in D$, are given by:
\begin{equation*}\label{estLL}
\hat m_{j, \bm{H}}(\bm{x};1)=\left\{\begin{array}{lc}\bm{e}_1^\top(\bm{X}_{\bm{x}}^\top\bm{W}_{\bm{x}}\bm{X}_{\bm{x}})^{-1}\bm{X}_{\bm{x}}^\top\bm{W}_{\bm{x}}\bm{S}&\text{if $j=1$},\\\\ \bm{e}_1^\top(\bm{X}_{\bm{x}}^\top\bm{W}_{\bm{x}}\bm{X}_{\bm{x}})^{-1}\bm{X}_{\bm{x}}^\top\bm{W}_{\bm{x}}\bm{C}&\text{if $j=2$},\end{array}\right.
\end{equation*}
where $\bm{e}_1$ is a $(d + 1) \times 1$ vector having 1 in the first entry and 0 in all other entries, $\bm{X}_{\bm{x}}$ is a $n\times(d+1)$ matrix having $(1, (\bm{X}_i-\bm{x})^\top)$ as its  $i$th row, $\bm{W}_{\bm{x}}=\mbox{diag}\{K_{\bm{H}}(\bm{X}_1-\bm{x}),\dots,K_{\bm{H}}(\bm{X}_n-\bm{x})\}$, $\bm{S}=(\sin(\Theta_1),\dots,\sin(\Theta_n))^\top$ and $\bm{C}=(\cos(\Theta_1),\dots,\cos(\Theta_n))^\top$.

		  \cite{meilan2019nonparametric} derived
		the AMSE of the circular regression estimator given in (\ref{est}), for independent data. In the present work, this estimator is studied for spatially correlated data, for $p=0$ {(corresponding to a NW-type estimator)} and $p=1$ {(providing a LL-type estimator)}.

%----------------------------------------------------------------%		
\section{{Theoretical results}}\label{sec:est}
%----------------------------------------------------------------%		
		{Some asymptotic conditional properties of  {the} estimator (\ref{est}), with polynomial degrees $p=0$  and $p=1$, are derived in Section \ref{sec:asy_results}. An asymptotically optimal local bandwidth matrix for $\hat{m}_{\bm{H}}(\bm{x};p)$, with $p=0$  and $p=1$, is also provided {in this section, whereas a suitably adapted cross-validation criterion is proposed in Section \ref{sec:smoothing}.}}

	\subsection{Asymptotic results}	\label{sec:asy_results}

	Asymptotic properties of  $\hat m_{j, {\bm{H}}}({\bm{x}};p)$, for $j=1,2$ and $p=0,1$, can be obtained using some results given in \cite{liu2001kernel}. The following assumptions on the design, the kernel function and the bandwidth matrix are needed to derive the asymptotic bias, the asymptotic variance of estimator $\hat m_{j, {\bm{H}}}({\bm{x}};p)$,  $j=1,2$, and the asymptotic covariance between $\hat m_{1, {\bm{H}}}({\bm{x}};p)$ and $\hat m_{2, {\bm{H}}}({\bm{x}};p)$, as well the asymptotic bias and variance of $\hat m_{ {\bm{H}}}({\bm{x}};p)$, $p=0,1$.

		\begin{enumerate}[{(A}1)]
			\item The design density $f$ is continuously differentiable at ${\bm{x}}\in D$, and satisfies  $f({\bm{x}})>0$. Moreover,  $s_j^2({\bm{x}})>0$, and $s_j^2$ and all second-order derivatives  of the regression functions $m_j$, for $j=1,2$,  are continuous at ${\bm{x}}$.
		\item  The kernel $K$ is a spherically symmetric density function, twice continuously differentiable, with compact support (for simplicity with a nonzero value only if $\norm{{\bm{u}}}\le 1$). Moreover, $\int {\bm{u}\bm{u}^\top} K({\bm{u}}){\textrm{d}}\bm{u}={\mu_2}{\bm{I}}_d$, where ${{\mu_2}=\int u^2_iK({\bm{u}}){\textrm{d}}\bm{u}\neq 0}$, for all $i=1,\dots,d$, and ${\bm{I}}_d$ denotes the $d\times d$ identity matrix. {Then, $\mu_2$ is the second-order moment of the multivariate kernel $K$.} It is also assumed that ${\nu_0}=\int K^2({\bm{u}}){\textrm{d}}\bm{u}<\infty$. 
			\item $K$ is Lipschitz continuous. That is, there exists a constant $\mathfrak{L}>0$, such that, 
			$$\abs{K({\bm{X}}_1)-K({\bm{X}}_2)}\le \mathfrak{L}\norm{{\bm{X}}_1-{\bm{X}}_2}, \quad \forall {\bm{X}}_1,{\bm{X}}_2\in D.$$
			\item The bandwidth matrix ${\bm{H}}$ is symmetric and positive definite, with ${\bm{H}}\to 0$ and  $n\abs{{\bm{H}}}\lambda^d_{\min}({\bm{H}})\to\infty$, when $n\to\infty$. 
			The ratio $\lambda_{{\max}}({\bm{H}})/\lambda_{{\min}}({\bm{H}})$ is bounded above
			\item For the correlation functions ${\rho_{k,n}} $, $k=1,2,3$, {in (\ref{cor1}), (\ref{cor2}) and (\ref{cor3}), respectively,} there exist {constants} $\rho_{{\textrm{M}}_{k}}$ and $\rho_{{\textrm{c}}_k}$, such that,		
		$  n\int \abs{{\rho_{k,n}}({\bm{x}})}{\textrm{d}}\bm{x}<\rho_{{\textrm{M}}_{k}}$ and $\lim_{n\to\infty}n\int {{\rho_{k,n}}({\bm{x}})}{\textrm{d}}\bm{x}=\rho_{{\textrm{c}}_k}.$
			Moreover, for any sequence $\epsilon_n>0$ satisfying $n^{1/2}\epsilon_n\to\infty$,
			$$n\int_{\norm{{\bm{x}}}\ge\epsilon_n} \abs{{\rho_{k,n}}({\bm{x}})}{\textrm{d}}\bm{x}\to 0\quad \text{as}\quad n\to\infty.$$
		\end{enumerate}
		
		In assumption (A4), ${\bm{H}}\to 0$ means that every entry of ${\bm{H}}$ goes to $0$. This condition is equivalent to $\lambda_{{\max}}({\bm{H}})\to 0$ due to the symmetry and positive definiteness of  ${\bm{H}}$. Further, $\abs{{\bm{H}}}=O[\lambda_{{\max}}^d({\bm{H}})]$, because $\abs{{\bm{H}}}$ is equal to the product of all eigenvalues of ${\bm{H}}$. Assumption (A5) implies that the correlation functions ${\rho_{k,n}}$, for { $k=1,2,3$,} depend on $n$, and  the integrals $\int \abs{{\rho_{k,n}}({\bm{x}})}{\textrm{d}}\bm{x}$,  { $k=1,2,3$,} should vanish  as $n\to\infty$. The vanishing speed should not be slower than $O(n^{-1})$. {For all  $k=1,2,3$,} this assumption also  entails that the integrals of $\abs{{\rho_{k,n}}({\bm{x}})}$  are essentially dominated by the values of ${\rho_{k,n}}({\bm{x}})$ near  the origin $\bm{0}$. Hence, the correlations
		are assumed to be short-ranged. This means that they decrease rapidly when the distance between two observations increases,  as $n\to\infty$.  Two examples of correlation functions that satisfy the conditions of assumption (A5) are the exponential model
		$${\rho_{n}}({\bm{x}})={\rm exp}(-a n\norm{{\bm{x}}}),$$
		and the rational quadratic model
		$${\rho_{n}}({\bm{x}})=\dfrac{1}{1+a(n\norm{{\bm{x}}})^2},$$
$a$ being a positive constant in both cases  \citep[see][]{cressie1993statistics}.

The asymptotic conditional bias and variance of the circular regression estimator $\hat{m}_{{\bm{H}}}({\bm{x}};p)$, for $p=0,1$, given in (\ref{est}), can be derived  by using the asymptotic bias and variance of  estimators $\hat m_{j, {\bm{H}}}({\bm{x}};p)$, $j= 1, 2$, as well as the asymptotic covariance between $\hat{m}_{1,{\bm{H}}}({\bm{x}};p)$ and $\hat{m}_{2,{\bm{H}}}({\bm{x}};p)$, for $p=0,1$. The asymptotic bias of estimators $\hat{m}_{j,{\bm{H}}}({\bm{x}};p)$, $j=1,2$,  assuming models (\ref{model1}) and (\ref{model2}), is the same as that obtained considering independent data, derived in \cite{hardle_muller} and \cite{ruppert1994multivariate}, for $p=0$ and $p=1$, respectively. On the other hand, their asymptotic variances and the covariances between $\hat{m}_{1,{\bm{H}}}({\bm{x}};p)$ and $\hat{m}_{2,{\bm{H}}}({\bm{x}};p)$, for $p=0,1$, (considering assumption (A5)) can be deduced from the results obtained in \cite{liu2001kernel}. These expressions are given in the final Appendix. Note that although assumption (A5) establishes conditions {on the correlations in (\ref{cor1}), (\ref{cor2}) and (\ref{cor3}), for the sine and cosine process, and the cross-correlation,} directly derived from model (\ref{model}), using the sine and cosine addition formulas, it is straightforward to obtain equations relating these covariances with those coming from models (\ref{model1}) and (\ref{model2}). For further details on the relation between $C_{n,3}$ and {$\rho_{k,n}$, $k=1,2,3,$ in (\ref{cor1}), (\ref{cor2}) and (\ref{cor3}),} see the final Appendix. Similar equations can be also obtained for $C_{n,1}$ and $C_{n,2}$. As in the Euclidean setting, the asymptotic bias of $\hat{m}_{{\bm{H}}}({\bm{x}};p)$, $p=0,1$, is the same for dependent and for independent data \citep{meilan2019nonparametric}. However, the asymptotic conditional variance of these estimators depends on the spatial correlation. Considering an interior point in  the support of $f$,  the NW- and  LL-type estimators of $m$ have the same asymptotic variance. The following theorem provides this result. Its proof is included in the  Appendix.
		
		\begin{theorem}\label{teoNW}
		Given a sample $\{({\bm{X}}_i,\Theta_i)\}_{i=1}^n$ on $D\times \mathbb{T}$ generated from model $(\ref{model})$ and assuming that conditions \textnormal{(A1)--(A5)} hold, the asymptotic conditional  variance of the estimator $\hat{m}_{{\bm{H}}}({\bm{x}};p)$, for $p=0,1$,  at  a fixed interior point ${\bm{x}}$ in the support of $f$, is given by:		
		\begin{eqnarray}
				{\mathbb{V}\rm ar}[\hat{m}_{{\bm{H}}}({\bm{x}};p)\mid{\bm{X}}_1,\dots,{\bm{X}}_n]&=&\dfrac{{\nu_0}\sigma^2_1[1+f({\bm{x}})\rho_{{\textrm{c}}_1}]}{n\abs{{\bm{H}}}\ell^2({\bm{x}})f({\bm{x}})}\nonumber\\&&+o_{\mathbb{P}}\bigg(\frac{1}{n \abs{{\bm{H}}}}\bigg),\label{variance}	\end{eqnarray}	
		where $\ell({\bm{x}})={\mathbb{E}}[\cos (\varepsilon)\mid{\bm{X}}={\bm{x}}]$.
		\end{theorem}
%		\begin{proof}See Appendix.
%		\end{proof}
		
		\begin{rem}  \normalfont Notice that  the expression of the asymptotic conditional variance of estimator $\hat{m}_{{\bm{H}}}({\bm{x}};p)$, for $p=0,1$, has a similar structure to that obtained for the NW and LL  estimators in a regression model with Euclidean response and spatially correlated errors. For independent data, it follows that $\rho_{{\textrm{c}}_1}=0$ in Theorem \ref{teoNW}, and, consequently, the asymptotic conditional variance of both estimators coincides with the expression obtained for  independent data in  \cite{meilan2019nonparametric}. 
		\end{rem}

 The AMSE of $\hat{m}_{{\bm{H}}}({\bm{x}};0)$, defined as the sum of the square of the leading term of the bias \citep{meilan2019nonparametric} and the leading term of the variance (\ref{variance}), is given by:
\begin{eqnarray}\label{AMSENW}
\mbox{AMSE}[\hat{m}_{{\bm{H}}}({\bm{x}};0)]&=&\Bigg\{\dfrac{1}{2}{{\mu_2}}{\rm tr}[{\bm{H}}^2\bm{\mathcal{H}}_{m}({\bm{x}})]\nonumber\\&&+ \dfrac{{\mu_2}}{\ell({\bm{x}})f({\bm{x}})}\bm{\nabla}{^\top} m({\bm{x}}){\bm{H}}^2\bm{\nabla}  (\ell f)({\bm{x}})\Bigg\}^2\nonumber\\&&+ \dfrac{{\nu_0}\sigma^2_1[1+f({\bm{x}})\rho_{{\textrm{c}}_1}]}{n\abs{{\bm{H}}}\ell^2({\bm{x}})f({\bm{x}})}\nonumber\\\nonumber&=&\dfrac{1}{4}{\mu^2_2}{\rm tr}^2[{\bm{H}}^2\mathcal{B}({\bm{x}})]\nonumber\\&&+ \dfrac{{\nu_0}\sigma^2_1[1+f({\bm{x}})\rho_{{\textrm{c}}_1}]}{n\abs{{\bm{H}}}\ell^2({\bm{x}})f({\bm{x}})},\end{eqnarray}
with
\begin{eqnarray*}\mathcal{B}({\bm{x}})&=&\dfrac{1}{\ell({\bm{x}})f({\bm{x}})}[\bm{\nabla} (\ell f)({\bm{x}})\bm{\nabla} {^\top} m({\bm{x}})\\&&+\bm{\nabla} m({\bm{x}})\bm{\nabla} {^\top} (\ell f)({\bm{x}})]+ \bm{\mathcal{H}}_{m}({\bm{x}}).\end{eqnarray*} 

An asymptotically optimal local bandwidth matrix, ${\bm{H}}_{ \text{opt}}({\bm{x}};0)$, for $\hat{m}_{{\bm{H}}}({\bm{x}};0)$ can be directly derived minimizing equation (\ref{AMSENW}) with respect to ${\bm{H}}$.
{
Using Proposition 2.6 of \cite{liu2001kernel}, it can be obtained that this optimal local bandwidth is:
}
\begin{eqnarray}\label{ob1NW}
{\bm{H}}_{ \text{opt}}({\bm{x}};0)=h^*({\bm{x}})\cdot[\tilde{\mathcal{B}}({\bm{x}})]^{-1/2},\end{eqnarray}
where
$$h^*({\bm{x}})=\bigg\{\dfrac{{\nu_0}\abs{\tilde{\mathcal{B}}({\bm{x}})}^{1/2}\sigma^2_1[1+f({\bm{x}})\rho_{{\textrm{c}}_1}]}{nd{\mu^2_2}\ell^2({\bm{x}})f({\bm{x}})}\bigg\}^{1/{d+4}},$$
and
$$\tilde{\mathcal{B}}({\bm{x}})= \bigg\{ \begin{array}{lcc}
\mathcal{B}({\bm{x}}) &   \text{ if }   & \mathcal{B}({\bm{x}}) \text{ is positive definite,} \\
-\mathcal{B}({\bm{x}}) & \text{if} & \mathcal{B}({\bm{x}}) \text{ is negative definite.} \\
\end{array}\big.$$

The matrix  $\tilde{\mathcal{B}}({\bm{x}})$  determines the shape and the orientation in the $d$-dimensional space of the covariate region which is used to compute the local estimates. Such data regions are ellipsoids in $\mathbb R^d$, being the magnitude of the axes controlled by $\tilde{\mathcal{B}}({\bm{x}})$. 	Similarly, an asymptotically optimal local bandwidth can be also obtained
for the LL-type estimator. In this case, the AMSE of $\hat{m}_{{\bm{H}}}({\bm{x}};1)$ is given by: \begin{eqnarray*}\label{AMSELL}
\mbox{AMSE}[\hat{m}_{{\bm{H}}}({\bm{x}};1)]\nonumber&=&\dfrac{1}{4}{\mu^2_2}{\rm tr}^2[{\bm{H}}^2\mathcal{G}({\bm{x}})]\nonumber\\&&+ \dfrac{{\nu_0}\sigma^2_1[1+f({\bm{x}})\rho_{{\textrm{c}}_1}]}{n\abs{{\bm{H}}}\ell^2({\bm{x}})f({\bm{x}})},\end{eqnarray*}
with
$ \mathcal{G}({\bm{x}})=\ell^{-1}({\bm{x}})[\bm{\nabla} \ell ({\bm{x}})\bm{\nabla} {^\top} m({\bm{x}})+\bm{\nabla} m({\bm{x}})\bm{\nabla} {^\top} \ell ({\bm{x}})]+\bm{\mathcal{H}}_{m}({\bm{x}})$.
 Consequently, the bandwidth matrix which minimizes this expression coincides with (\ref{ob1NW}), but using  $ \mathcal{G}({\bm{x}})$ instead of $ \mathcal{B}({\bm{x}})$.

Local bandwidth matrices may be useful for estimating the trend at a
given point ${\bm{x}}\in D$, however, the nonparametric estimators computed with them may not be accurate enough for reconstructing the whole trend. An asymptotically optimal global bandwidth matrix can be selected minimizing the asymptotic mean integrated squared error (AMISE). Unfortunately, there is not a closed form solution for this optimization problem. Moreover, optimal bandwidth matrices, depending on unknown quantities, cannot be used for practical purposes. Practical bandwidth selection techniques, based on cross-validation methods, are considered in what follows.

%--------------------------------------------------------------------%
\subsection{{Cross-validation {bandwidth selection methods}}}\label{sec:smoothing}
%--------------------------------------------------------------------%
	 A first approach to select  the bandwidth ${\bm{H}}$ {for $\hat{m}_{{\bm{H}}}({\bm{x}};p)$, $p=0,1$,} consists on minimizing the cross-validation function:
	$${\rm CV}({\bm{H}})=\sum_{i=1}^n \big\{  1-\cos\big[\Theta_i-\hat{m}_{{\bm{H}}}^{(i)}({\bm{X}}_i;p)\big]\big\},$$ 
	where $\hat{m}_{{\bm{H}}}^{(i)}({\bm{X}}_i;p)$ is  the  estimator  computed using all observations except $({\bm{X}}_i,\Theta_i)$ and evaluated at ${\bm{X}}_i$. The CV criterion, as well as other smoothing parameter selection methods in nonparametric regression, should not be directly used  for selecting the bandwidth when working with dependent data, given that its
	expectation is severely affected by the correlation  \citep{opsomer2001nonparametric,liu2001kernel}. 
	
	In our setting, the CV criterion should be modified in order to account for the effect of the spatial correlation. With this issue in mind, we propose a modified cross-validation (MCV) criterion, which selects the bandwidth matrix ${\bm{H}}$ that minimizes the function:
		$${\rm MCV}({\bm{H}})=\sum_{i=1}^n \big\{  1-\cos\big[\Theta_i-\hat{m}_{{\bm{H}}}^{N(i)}({\bm{X}}_i;p)\big]\big\},$$ 
	where $\hat{m}_{{\bm{H}}}^{N(i)}({\bm{X}}_i;p)$ denotes the estimator  computed using all observations except  those  located  within a neighborhood of ${\bm{X}}_i$, $N(i)$, and evaluated at ${\bm{X}}_i$.  For applying this criterion, the size of the neighborhood $N{(i)}$ must be selected. {For simplicity, we consider} the MCV criterion when 
	$N{(i)}=\{{\bm{X}}_j: \norm{{\bm{X}}_j-{\bm{X}}_i}\le {l}\}$. 
For $d=2$, the neighborhood $N{(i)}$ consists of observations within the circle centered at ${\bm{X}}_i$ and  radius 
{$l$}.  If there is a strong spatial correlation,   more observations should be omitted in the bandwidth selection procedure, and consequently,  the value of $l$ for constructing $N{(i)}$ should be larger. The use of the CV and MCV criteria to select the bandwidth matrix is explored through simulations in the  following section.

%-----------------------------------------------------------------------%
\section{Simulation study}\label{sec:sim}
%-----------------------------------------------------------------------%
					\begin{figure*}
			\centering
			\includegraphics[width=1\textwidth]{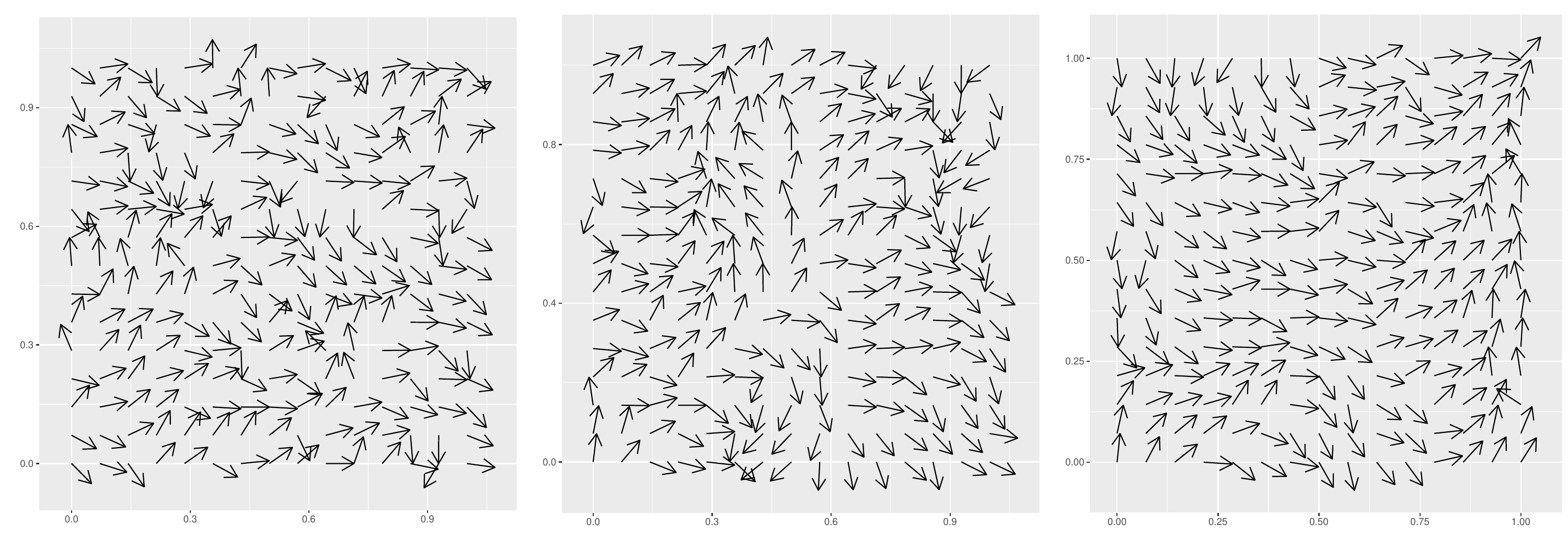}
			\caption{Simulated samples of a   wrapped Gaussian spatial process on a $15\times 15$ grid with exponential correlation, being $a_\textrm{e}=0.1$  (left), $a_\textrm{e}=0.3$ (center) and $a_\textrm{e}=0.6$ (right), for $\mu=0$ and $\sigma=1$ in (\ref{processdecomposedW}) and (\ref{expo_w}).}
			\label{figure:sim_wrap}
		\end{figure*}

		 \begin{table*}[h!]
	\centering
	\caption{{Results obtained when the errors in model (\ref{model}) are simulated from wrapped Gaussian spatial processes. Average (over $500$ replicates) of the CASE given in (\ref{CASE}), for the  regression function $r_1$, using the  NW type estimator. Bandwidth matrix is selected by minimizing CV ($\bm{H}_{\rm CV}$), MCV ($\bm{H}_{{\rm MCV}_{1}}$, $\bm{H}_{{\rm MCV}_{2}}$, $\bm{H}_{{\rm MCV}_{3}}$) and CASE ($\bm{H}_{{\rm CASE}}$) as a benchmark.}}
	\begin{tabular}{ccccccc}
	&&	&&	 NW&& \\
	\cline{3-7} 
	$a_\textrm{e}$&$n$&${\bm{H}}_{\rm CV}$&${\bm{H}}_{{\rm MCV}_{1}}$&${\bm{H}}_{{\rm MCV}_{2}}$&${\bm{H}}_{{\rm MCV}_{3}}$&${\bm{H}}_{{\rm CASE}}$\\	
	 \hline
0.1 & 100 & 0.2087 & 0.0902 & 0.0721 & 0.0609 & 0.0387   \\ 
& 225 & 0.2880 & 0.1291 & 0.0768 & 0.0602 & 0.0365    \\ 
& 400 & 0.2932 & 0.1195 & 0.0702 & 0.0585 & 0.0359    \\ \cline{2-7}
0.3 & 100 & 0.2852 & 0.1752 & 0.1342 & 0.0803 & 0.0529    \\ 
& 225 & 0.3080 & 0.2054 & 0.1500 & 0.0788 & 0.0520  \\ 
& 400 & 0.2764 & 0.1967 & 0.1351 & 0.0778 & 0.0497    \\ \cline{2-7}
0.6 & 100 & 0.2316 & 0.1591 & 0.1316 & 0.0806 & 0.0677    \\ 
& 225 & 0.2417 & 0.1775 & 0.1455 & 0.0798 & 0.0620    \\ 
& 400 & 0.2177 & 0.1701 & 0.1325 & 0.0778 & 0.0569  \\ 
\hline
\end{tabular}
						\label{table:simus_wrap_R1b}
		\end{table*}
		
		\begin{table*}[h!]
	\centering
	\caption{{Results obtained when the errors in model (\ref{model}) are simulated from wrapped Gaussian spatial processes. Average (over $500$ replicates) of the CASE given in (\ref{CASE}), for the  regression function $r_1$, using the  LL type estimator. Bandwidth matrix is selected by minimizing CV ($\bm{H}_{\rm CV}$), MCV ($\bm{H}_{{\rm MCV}_{1}}$, $\bm{H}_{{\rm MCV}_{2}}$, $\bm{H}_{{\rm MCV}_{3}}$) and CASE ($\bm{H}_{{\rm CASE}}$) as a benchmark.}}
	\begin{tabular}{ccccccc}
	&&	&&	 LL&& \\
	\cline{3-7} 
	$a_\textrm{e}$&$n$&${\bm{H}}_{\rm CV}$&${\bm{H}}_{{\rm MCV}_{1}}$&${\bm{H}}_{{\rm MCV}_{2}}$&${\bm{H}}_{{\rm MCV}_{3}}$&${\bm{H}}_{{\rm CASE}}$\\	
	 \hline
0.1 & 100 & 0.1818 & 0.0933 & 0.0839 & 0.0672 & 0.0550   \\ 
& 225 & 0.2649 & 0.1190 & 0.0782 & 0.0667 & 0.0532    \\ 
& 400 & 0.2920 & 0.1114 & 0.0771 & 0.0667 & 0.0518    \\ \cline{2-7}
0.3 & 100 & 0.2499 & 0.1651 & 0.1474 & 0.1143 & 0.1062    \\ 
& 225 & 0.2979 & 0.2026 & 0.1546 & 0.1196 & 0.1053  \\ 
& 400 & 0.2785 & 0.1965 & 0.1495 & 0.1168 & 0.1019    \\ \cline{2-7}
0.6 & 100 & 0.2117 & 0.1520 & 0.1392 & 0.1146 & 0.1097    \\ 
& 225 & 0.2361 & 0.1783 & 0.1488 & 0.1212 & 0.1093    \\ 
& 400 & 0.2192 & 0.1725 & 0.1413 & 0.1171 & 0.1074  \\ 
\hline
\end{tabular}
						\label{table:simus_wrap_R1c}
		\end{table*}

		\begin{table*}[h!]
	\centering
	\caption{{Results obtained when the errors in model (\ref{model}) are simulated from wrapped Gaussian spatial processes. Average (over $500$ replicates) of the CASE given in (\ref{CASE}), for the  regression function $r_2$, using the  NW  type estimator. Bandwidth matrix is selected by minimizing CV ($\bm{H}_{\rm CV}$), MCV ($\bm{H}_{{\rm MCV}_{1}}$, $\bm{H}_{{\rm MCV}_{2}}$, $\bm{H}_{{\rm MCV}_{3}}$) and CASE ($\bm{H}_{{\rm CASE}}$) as a benchmark.}}
		\begin{tabular}{ccccccc}
		&&	&&	 NW&& \\
		\cline{3-7} 
		$a_\textrm{e}$&$n$&${\bm{H}}_{\rm CV}$&${\bm{H}}_{{\rm MCV}_{1}}$&${\bm{H}}_{{\rm MCV}_{2}}$&${\bm{H}}_{{\rm MCV}_{3}}$&${\bm{H}}_{{\rm CASE}}$\\	
 \hline
0.1 & 100 & 0.2231 & 0.1331 & 0.1233 & 0.1282 & 0.0871  \\ 
& 225 & 0.3044 & 0.1475 & 0.1120 & 0.1122 & 0.0792  \\ 
& 400 & 0.2941 & 0.1369 & 0.1008 & 0.1051 & 0.0748  \\ \cline{2-7}
0.3 & 100 & 0.1927 & 0.1740 & 0.1622 & 0.1531 & 0.1277  \\ 
& 225 & 0.2053 & 0.1879 & 0.1620 & 0.1519 & 0.1030  \\ 
& 400 & 0.2154 & 0.1936 & 0.1529 & 0.1414 & 0.1017  \\ \cline{2-7}
0.6 & 100 & 0.2392 & 0.1704 & 0.1605 & 0.1561 & 0.1440  \\ 
& 225 & 0.1891 & 0.1765 & 0.1729 & 0.1553 & 0.1196  \\ 
& 400 & 0.1947 & 0.1687 & 0.1624 & 0.1467 & 0.1063  \\ 
\hline

 	\end{tabular}
			\label{table:simus_wrap_R2b}
		\end{table*}

	\begin{table*}[h!]
	\centering
	\caption{{Results obtained when the errors in model (\ref{model}) are simulated from wrapped Gaussian spatial processes. Average (over $500$ replicates) of the CASE given in (\ref{CASE}), for the  regression function $r_2$, using the  LL  type estimator. Bandwidth matrix is selected by minimizing CV ($\bm{H}_{\rm CV}$), MCV ($\bm{H}_{{\rm MCV}_{1}}$, $\bm{H}_{{\rm MCV}_{2}}$, $\bm{H}_{{\rm MCV}_{3}}$) and CASE ($\bm{H}_{{\rm CASE}}$) as a benchmark.}}
		\begin{tabular}{ccccccc}
		&&	&&	 LL&& \\
		\cline{3-7} 
		$a_\textrm{e}$&$n$&${\bm{H}}_{\rm CV}$&${\bm{H}}_{{\rm MCV}_{1}}$&${\bm{H}}_{{\rm MCV}_{2}}$&${\bm{H}}_{{\rm MCV}_{3}}$&${\bm{H}}_{{\rm CASE}}$\\	
 \hline
0.1 & 100 & 0.1804 & 0.1163 & 0.1092 & 0.1114 & 0.078  \\ 
& 225 & 0.2574 & 0.1264 & 0.0976 & 0.0980 & 0.0704  \\ 
& 400 & 0.2916 & 0.1160 & 0.0886 & 0.0973 & 0.0668  \\ \cline{2-7}
0.3 & 100 &  0.1991 & 0.1688 & 0.1627 & 0.1526 & 0.1299 \\ 
& 225 & 0.2090 & 0.1869 & 0.1563 & 0.1518 & 0.1277  \\ 
& 400 & 0.2189 & 0.1875 & 0.1505 & 0.1432 & 0.1208   \\ \cline{2-7}
0.6 & 100 & 0.2072 & 0.1605 & 0.1602 & 0.1597 & 0.1348  \\ 
& 225 & 0.1903 & 0.1717 & 0.1669 & 0.1594 & 0.1343  \\ 
& 400 & 0.1962 & 0.1653 & 0.1608 & 0.1509 & 0.1227  \\ 
\hline

 	\end{tabular}
			\label{table:simus_wrap_R2c}
		\end{table*}

{In this section, the performance of the proposed estimators and the cross-validation bandwidth selection criteria are analyzed in a simulation study for $d=2$. Section \ref{sec:sim_gen} contains a detailed description of the {different simulation scenarios.} In Sections \ref{sec:wra} and \ref{sec:pro}, we  briefly describe the procedures employed to generate  wrapped and projected Gaussian spatial  errors in model (\ref{model}), respectively. The practical performance of the proposed circular regression estimator { in (\ref{est})} is analyzed considering these {previous} error simulation approaches.}

\subsection{{General aspects}}\label{sec:sim_gen}

 Assuming regression model (\ref{model}), 500 samples of size $n$ ($n=100, 225, 400$) are generated, considering {the spatial locations} ${\bm{X}}=(X_{1},X_{2})$ on a bidimensional regular grid in the unit square. Two different regression functions (shown in the left panels of Fig. \ref{C_figure:regression})  are considered:
\begin{eqnarray*}
r_1: m({\bm{X}})&=&{\mbox{arctan2}}(6{X}_{1}^5-2{X}_{1}^3-1,-2{X}_{2}^5-3{X}_{2}-1),\\
r_2: m({\bm{X}})&=&{\mbox{arccos}}({X}_{1}^5-1)+\dfrac{3}{2}{\mbox{arcsin}}({X}_{2}^3-{X}_{2}+1).
\end{eqnarray*}
		
Circular spatially correlated errors in model  (\ref{model}), for both regression functions, are generated from wrapped  \citep{jona2012spatial} and from projected Gaussian spatial processes \citep{wang2014modeling}. For each sample, the NW- and LL-type estimators of the circular regression function, given in (\ref{est}), are computed. In both cases,  a multiplicative triweight kernel is considered, while the bandwidth matrix ${{\bm{H}}}$ is selected by  using the ${\rm CV}$ and {MCV}  criteria. {Different values of the radius $l$ are considered in the MCV method.
Given that the covariates are located in unit square, we set $l(b)=\sqrt{2}b/10$, where $b=0$ would correspond with the CV method and $b=10$ would provide the maximum distance between two points in the unit square. After some tests, only three values of $b$ ($b=1,2, 3$) are considered. The corresponding CV and MCV bandwidths are
 denoted by $\bm{H}_{\rm CV}$ and $\bm{H}_{{\rm MCV}_b}$, $b=1,2,3$, respectively.} Taking into account the structure of the regression functions and in order to save computational time, the
		bandwidth matrix is restricted to be diagonal with possibly different elements. {The performance of the estimators and the bandwidth selectors is evaluated using the}  circular average squared error (CASE), defined as \begin{eqnarray}
	{\rm CASE}(\bm{H})=\frac{1}{n}\sum_{i=1}^n \left\{1- \cos\left[m(\bm{X}_i) - \hat{m}_{\bm{H}}(\bm{X}_i;p)\right]\right\},\hspace{0.45cm}\label{CASE}\end{eqnarray}
		for $p=0,1$,  as a comparative error measure \citep{kim2017multivariate}. Additionally, the diagonal optimal bandwidth matrix ${\bm{H}}_{{\rm CASE}}$  minimizing (\ref{CASE}), obtained by intensive search, is also computed. {Note that this bandwidth matrix can not be used in a practical situation where the true regression is unknown. For this reason, it can not be
considered as a criterion to select the bandwidth, but it is used to get a benchmark value for comparison.}
		
		{The computing time for running the whole procedure (simulate a sample, select the bandwidth matrix, compute the circular nonparametric estimator and evaluate the CASE) for just one of the 500 samples of size of 225 is around 2 seconds, no matter the bandwidth matrix selection method employed and regardless of the estimator (NW or LL) used. However, it should be noted that the computing time for obtaining a bandwidth matrix with the MCV criterion increases with $b$. In addition, the projection approach for the circular errors generation seems to be slightly more computationally expensive than the wrapping method.}

%----------------------------------------------------------------%		
\subsection{The wrapping approach}\label{sec:wra}
%----------------------------------------------------------------%
Given a collection of spatial coordinates, ${\bm{X}}_i$, with $i=1,\ldots,n$, a realization of a spatial circular (error) process $\{\varepsilon_i,\;i=1,\ldots,n\}$ can be obtained using the wrapping method, introduced by \cite{jona2012spatial}, for Gaussian spatial processes. In general, a wrapping approach consists on {wrapping}  a linear variable around the unit circle. In this case, its circular density function is easily obtained by wrapping the density function of the linear random variable.

So, consider a realization $\{Y_i=Y({\bm{X}}_i),\;i=1,\ldots,n\}$ from a real-valued Gaussian process, where each observation can be decomposed as:
\begin{equation}\label{processdecomposedW}
 Y_i=\mu+w_i, \quad i=1,\dots,n,
\end{equation} 
being $\mu=\mu({\bm{X}}_i)$ the mean and $w$ a zero mean Gaussian spatial process with ${\mathbb{C}\rm ov}(w_i,w_j\mid{\bm{X}}_i, {\bm{X}}_j)=\sigma^2\rho_{n}({\bm{X}}_i-{\bm{X}}_j)$. The variance of $w$ is denoted by $\sigma^2$ and $\rho_n$ is a continuous stationary correlation function satisfying $\rho_n(\bm{0})=1$, $\rho_n({\bm{x}})=\rho_n(-{\bm{x}})$, and $\abs{\rho_n({\bm{x}})}\le1$, $\forall {\bm{x}}$.  Then, a realization of a wrapped Gaussian spatial process $\{\varepsilon_i,\;i=1,\ldots,n\}$, linked to the spatial coordinates ${\bm{X}}_i$, with $i=1,\ldots,n$, is obtained as:
 $$\varepsilon_i=Y_i({\rm \texttt{mod}}\,2\pi), \quad i=1,\dots,n. $$
 
Note that this realization can be written in vector form as $\bm{\varepsilon}=(\varepsilon_1,\ldots,\varepsilon_n){^\top}$, with mean direction vector $\mu \bm{1}_n$, being $\bm{1}_n$ a $n\times  1$ vector with every entry equal to 1, and covariance matrix  $\sigma^2 {\bm{R}}_n$, where ${\bm{R}}_n(i,j)=\rho_n({\bm{X}}_i-{\bm{X}}_j)$ is the $(i,j)$-entry of the correlation matrix ${\bm{R}}_n$. In the simulation study, the unwrapped Gaussian spatial process to generate the errors is obtained assuming model (\ref{processdecomposedW}), with constant mean, $\mu$, equal to zero and exponential covariance structure
\begin{equation}
{\mathbb{C}\rm ov}({w}_i,w_j\mid{\bm{X}}_i, {\bm{X}}_j)=   
\sigma^2[\exp(-\norm{{\bm{X}}_i-{\bm{X}}_j}/a_\textrm{e})],
\label{expo_w}
\end{equation}where  $a_\textrm{e}$ is the practical
range. This exponential spatial correlation function and the circular correlation of the corresponding wrapped Gaussian spatial process were compared by \citet[][Fig. 4]{jona2012spatial}, obtaining very similar shapes for both correlations. Notice that although the vector of circular variables  $\bm{\varepsilon}$ has {almost} zero mean direction,  to properly apply the estimation procedure in practice, $\bm{\varepsilon}$ must be centered. In the simulation experiments,
the value of the variance $\sigma^2$ in (\ref{expo_w}) is set equal to one, and different values of the parameter $a_\textrm{e}$ are considered: $a_\textrm{e}=0.1$ (weak correlation), $a_\textrm{e}=0.3$ (medium correlation) and $a_\textrm{e}=0.6$ (strong correlation). Fixing the values of $\mu=0$ and $\sigma=1$,  the effect of the range parameter $a_\textrm{e}$ on a realization {(on a $15\times 15$ grid)} of the wrapped circular spatial process can be seen in Fig. \ref{figure:sim_wrap}. Larger values of the range {$a_\textrm{e}$} yield a {smoother}  pattern. 
 
For  the regression function $r_1$, Tables \ref{table:simus_wrap_R1b} and \ref{table:simus_wrap_R1c}  show the average, over 500 replicates, of the CASE given in (\ref{CASE}) considering the bandwidths selected by the  ${\rm CV}$ and   {MCV} 
methods, and the minimum value of ${\rm CASE}({\bm{H}})$, i.e., ${\rm CASE}({\bm{H}}_{{\rm CASE}})$, which can be viewed as a benchmark, for the NW- and LL- type estimators. Note that the optimal error increases as the dependence range becomes larger. It can be observed  the poor behavior of the ${\rm CV}$ bandwidth, providing average values of the ${\rm CASE}$ far from the optimal value, not even decreasing for large sample sizes. In general, the ${\rm MCV}$ criterion  appears to provide a significant improvement over the ${\rm CV}$ one when correlation is present.  It can be observed that ${\bm{H}}_{{\rm MCV}_{3}}$ provides good results for all cases, decreasing the error as $n$ gets larger. For stronger dependence (larger range values), this is the only selector that provides a reasonable behavior.  Similar conclusions can be {derived} when considering the regression function $r_2$. The corresponding results are displayed in Tables  \ref{table:simus_wrap_R2b} and \ref{table:simus_wrap_R2c}. Notice that when  $a_\textrm{e}=0.1$ (weak spatial correlation), the best behavior is observed when ${\bm{H}}_{{\rm MCV}_{2}}$ is employed. As expected, for larger values of the practical range $a_\textrm{e}$,  ${\bm{H}}_{{\rm MCV}_{3}}$ provides better results. Note that no major differences have been found if the NW- or LL-type estimators are employed.

 \begin{figure*}
	\centering
	\includegraphics[width=1\textwidth]{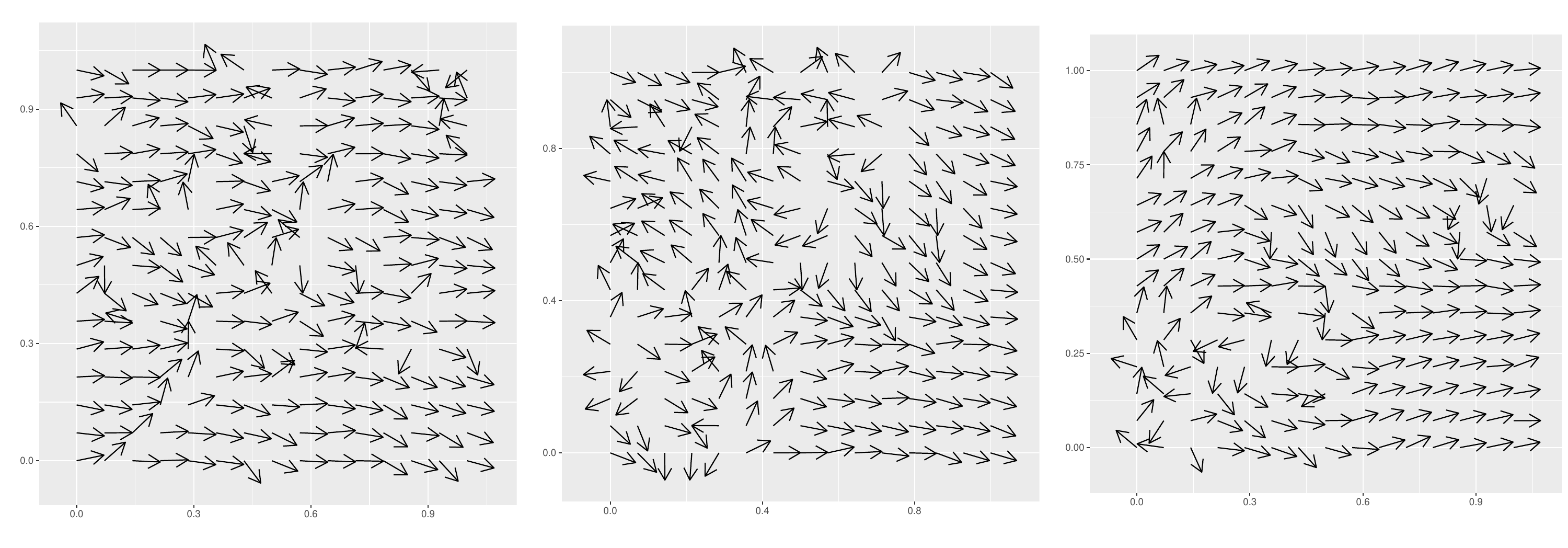}
	\caption{Simulated samples of a   projected Gaussian spatial process on a $15\times 15$ grid with exponential correlation being $a_\textrm{e}=0.1$  (left), $a_\textrm{e}=0.3$ (center) and $a_\textrm{e}=0.6$ (right), for $\bm{\mu}=(1,1){^\top}$, $\sigma=1$ and $\tau=0.9$ in (\ref{processdecomposedp}) and (\ref{expo_p}).}
	\label{figure:sim_proj}
\end{figure*}

\begin{table*}[h!]
	\centering
	\caption{{Results obtained when the errors in model (\ref{model}) are simulated from projected Gaussian spatial processes. Average (over $500$ replicates) of the CASE given in (\ref{CASE}), for the  regression function $r_1$, using the  NW type estimator. Bandwidth matrix is selected by minimizing CV ($\bm{H}_{\rm CV}$), MCV ($\bm{H}_{{\rm MCV}_{1}}$, $\bm{H}_{{\rm MCV}_{2}}$, $\bm{H}_{{\rm MCV}_{3}}$) and CASE ($\bm{H}_{{\rm CASE}}$) as a benchmark.}}
	\begin{tabular}{ccccccccccccc}
		&&	&&	 NW&& \\
		\cline{3-7} 
		$a_\textrm{e}$&$n$&${\bm{H}}_{\rm CV}$&${\bm{H}}_{{\rm MCV}_{1}}$&${\bm{H}}_{{\rm MCV}_{2}}$&${\bm{H}}_{{\rm MCV}_{3}}$&${\bm{H}}_{{\rm CASE}}$\\
		\hline
		0.1 & 100 & 0.1465 & 0.0629 & 0.0482 & 0.0452 & 0.0249  \\ 
		& 225 & 0.2207 & 0.0874 & 0.0417 & 0.0386 & 0.0216 \\ 
		& 400 & 0.2639 & 0.0722 & 0.0380 & 0.0385 & 0.0209 \\ 	\cline{2-7}
		0.3 & 100 & 0.2562 & 0.1746 & 0.1382 & 0.1142 & 0.0564  \\ 
		& 225 & 0.2235 & 0.1851 & 0.1248 & 0.1123 & 0.0558  \\ 
		& 400 & 0.2498 & 0.1987 & 0.1279 & 0.1109& 0.0474  \\ 	\cline{2-7}
		0.6 & 100 & 0.2368 & 0.1871 & 0.1567 & 0.1378 & 0.0620  \\ 
		& 225 & 0.2452 & 0.2126 & 0.1659 & 0.1371 & 0.0585  \\ 
		& 400 & 0.2424 & 0.1991 & 0.1600 & 0.1303 & 0.0525 \\ 
		\hline
	\end{tabular}

	\label{table:simus_proj_R1b}
\end{table*}

\begin{table*}[h!]
	\centering
	\caption{{Results obtained when the errors in model (\ref{model}) are simulated from projected Gaussian spatial processes. Average (over $500$ replicates) of the CASE given in (\ref{CASE}), for the  regression function $r_1$, using the  LL type estimator. Bandwidth matrix is selected by minimizing CV ($\bm{H}_{\rm CV}$), MCV ($\bm{H}_{{\rm MCV}_{1}}$, $\bm{H}_{{\rm MCV}_{2}}$, $\bm{H}_{{\rm MCV}_{3}}$) and CASE ($\bm{H}_{{\rm CASE}}$) as a benchmark.}}
	\begin{tabular}{ccccccccccccc}
		&&	&&	 LL&& \\
		\cline{3-7} 
		$a_\textrm{e}$&$n$&${\bm{H}}_{\rm CV}$&${\bm{H}}_{{\rm MCV}_{1}}$&${\bm{H}}_{{\rm MCV}_{2}}$&${\bm{H}}_{{\rm MCV}_{3}}$&${\bm{H}}_{{\rm CASE}}$\\
		\hline
		0.1 & 100 & 0.1376 & 0.0735 & 0.0667 & 0.0596 & 0.0381  \\ 
		& 225 & 0.1976 & 0.0891 & 0.0552 & 0.0566 & 0.0321 \\ 
		& 400 & 0.2430 & 0.0745 & 0.0497 & 0.0535 & 0.0301 \\ 	\cline{2-7}
		0.3 & 100 & 0.2452 & 0.1823 & 0.1658 & 0.1534 & 0.1205  \\ 
		& 225 & 0.2300 & 0.1948 & 0.1555 & 0.1498 & 0.1203  \\ 
		& 400 & 0.2524 & 0.1974 & 0.1554 & 0.1478 & 0.1125  \\ 	\cline{2-7}
		0.6 & 100 & 0.1988 & 0.1772 & 0.1703 & 0.1622 & 0.1376  \\ 
		& 225 & 0.2072 & 0.1927 & 0.1710 & 0.1612 & 0.1361  \\ 
		& 400 & 0.2051 & 0.1868 & 0.1637 & 0.1578 & 0.1300 \\ 
		\hline
	\end{tabular}

	\label{table:simus_proj_R1c}
\end{table*}

\begin{table*}[h!]
	\centering
		\caption{{Results obtained when the errors in model (\ref{model}) are simulated from projected Gaussian spatial processes. Average (over $500$ replicates) of the CASE given in (\ref{CASE}), for the  regression function $r_2$, using the NW type estimators. Bandwidth matrix is selected by minimizing CV ($\bm{H}_{\rm CV}$), MCV ($\bm{H}_{{\rm MCV}_{1}}$, $\bm{H}_{{\rm MCV}_{2}}$, $\bm{H}_{{\rm MCV}_{3}}$) and CASE ($\bm{H}_{{\rm CASE}}$) as a benchmark.}}
	\begin{tabular}{ccccccc}
		&&	&&	 NW&& \\
		\cline{3-7}
		$a_\textrm{e}$&$n$&${\bm{H}}_{\rm CV}$&${\bm{H}}_{{\rm MCV}_{1}}$&${\bm{H}}_{{\rm MCV}_{2}}$&${\bm{H}}_{{\rm MCV}_{3}}$&${\bm{H}}_{{\rm CASE}}$\\	
		\hline
		0.1 & 100 & 0.1764 & 0.1111 & 0.1055 & 0.1083 & 0.0813  \\ 
		& 225 & 0.2324 & 0.1177 & 0.0865 & 0.0901 & 0.0578  \\ 
		& 400 & 0.2689 & 0.1048 & 0.0772 & 0.0854 & 0.0516 \\ 
		\cline{2-7}
		0.3 & 100 & 0.2701 & 0.2132 & 0.1822 & 0.1752 & 0.1457  \\ 
		& 225 & 0.2890 & 0.2354 & 0.1919 & 0.1723 & 0.1043  \\ 
		& 400 & 0.2850 & 0.2223 & 0.1734 & 0.1583 & 0.0940  \\ 	\cline{2-7}
		0.6 & 100 & 0.2458 & 0.2068 & 0.1934 & 0.1847 & 0.1472  \\ 
		& 225 & 0.2558 & 0.2211 & 0.1953 & 0.1777 & 0.1010  \\ 
		& 400 & 0.2444 & 0.2082 & 0.1806 & 0.1679 & 0.0960  \\ 
		\hline
	\end{tabular}
	\label{table:simus_proj_R2b}
\end{table*}

\begin{table*}[h!]
	\centering
		\caption{{Results obtained when the errors in model (\ref{model}) are simulated from projected Gaussian spatial processes. Average (over $500$ replicates) of the CASE given in (\ref{CASE}), for the  regression function $r_2$, using the LL type estimators. Bandwidth matrix is selected by minimizing CV ($\bm{H}_{\rm CV}$), MCV ($\bm{H}_{{\rm MCV}_{1}}$, $\bm{H}_{{\rm MCV}_{2}}$, $\bm{H}_{{\rm MCV}_{3}}$) and CASE ($\bm{H}_{{\rm CASE}}$) as a benchmark.}}
	\begin{tabular}{ccccccc}
		&&	&&	 LL&& \\
		\cline{3-7}
		$a_\textrm{e}$&$n$&${\bm{H}}_{\rm CV}$&${\bm{H}}_{{\rm MCV}_{1}}$&${\bm{H}}_{{\rm MCV}_{2}}$&${\bm{H}}_{{\rm MCV}_{3}}$&${\bm{H}}_{{\rm CASE}}$\\	
		\hline
		0.1 & 100 & 0.1551 & 0.1023 & 0.0935 & 0.0957 & 0.0792  \\ 
		& 225 & 0.2023 & 0.1024 & 0.0769 & 0.0790 & 0.0513 \\ 
		& 400 & 0.2470 & 0.0909 & 0.0701 & 0.0717 & 0.0465 \\ 
		\cline{2-7}
		0.3 & 100 & 0.2510 & 0.1998 & 0.1832 & 0.1796 & 0.1621 \\ 
		& 225 &  0.2708 & 0.2268 & 0.1913 & 0.1749 & 0.1412  \\ 
		& 400 & 0.2773 & 0.2134 & 0.1769 & 0.1687 & 0.1326  \\ 	\cline{2-7}
		0.6 & 100 & 0.2344 & 0.1995 & 0.1900 & 0.1877 & 0.1849  \\ 
		& 225 & 0.2441 & 0.2143 & 0.1923 & 0.1856 & 0.1517  \\ 
		& 400 & 0.2403 & 0.2017 & 0.1801 & 0.1752 & 0.1439  \\ 
		\hline
	\end{tabular}
	\label{table:simus_proj_R2c}
\end{table*}

	\begin{figure*}[t]
	\centering
	\includegraphics[width=1\textwidth]{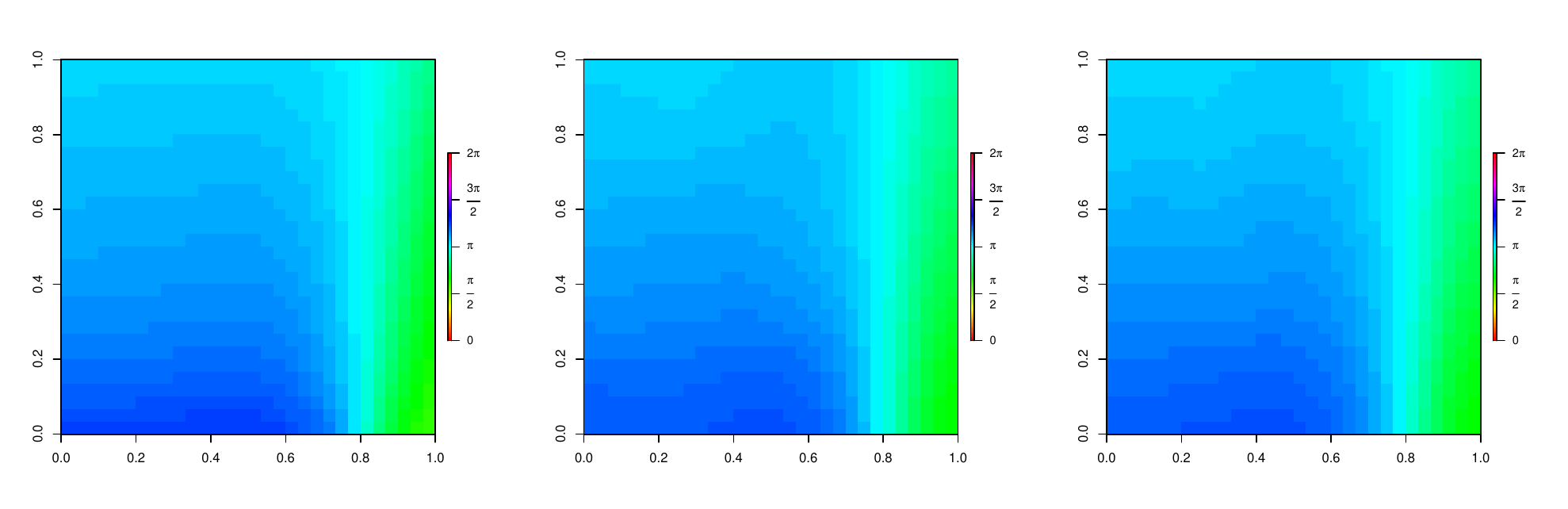}
	\includegraphics[width=1\textwidth]{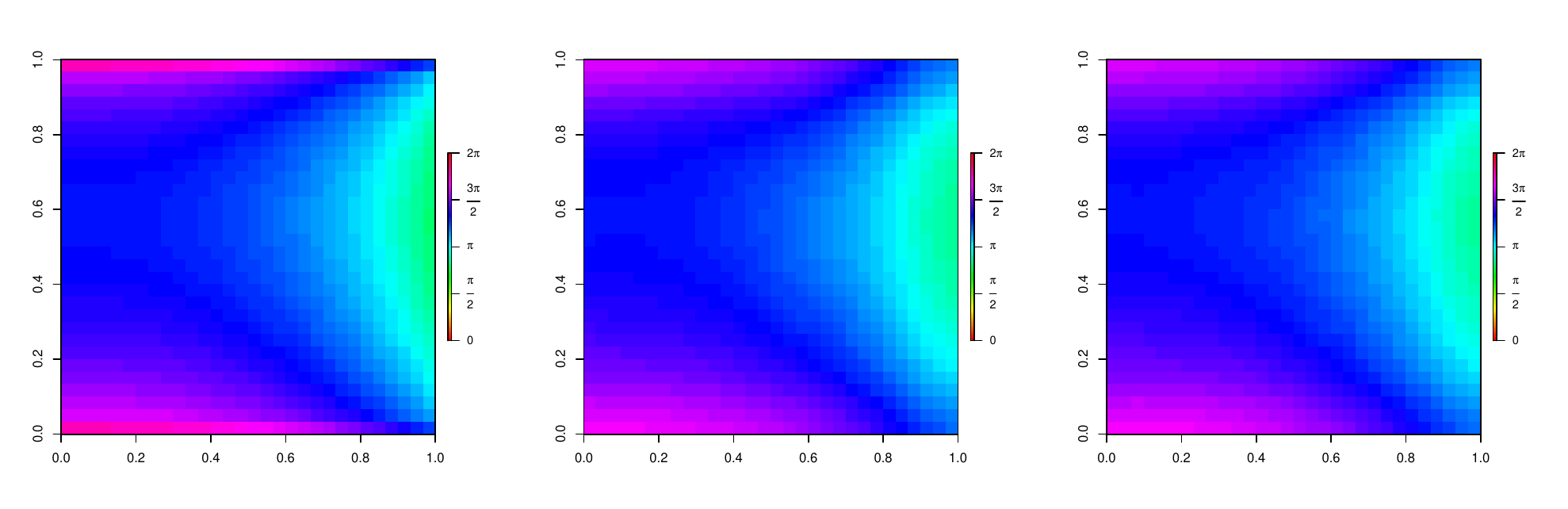}
	\caption{Theoretical regression function (left), jointly with the NW-type estimators using wrapped (center) and projected Gaussian spatial processes (right) to generate the errors in model (\ref{model}), for the regression function $r_1$ (top row) and  $r_2$ (bottom row).}
	\label{C_figure:regression}
\end{figure*}

%----------------------------------------------------------------%
\subsection{The projection approach}\label{sec:pro}	
%----------------------------------------------------------------%	
An alternative way of generating a realization from a circular spatial process is by considering a projection approach, as proposed by \cite{wang2014modeling}. A projected Gaussian spatial process is constructed from a bivariate Gaussian spatial process, and their correlation structures can be related, as shown in \cite{wang2014modeling}. This procedure allows to obtain samples of a projected spatial circular error as follows. First,  a bivariate Gaussian spatial process, ${\bm{Y}}$, observed at a collection of spatial coordinates ${\bm{X}}_i$, with $i=1,\ldots,n$, is considered. The observations  ${\bm{Y}}_{i}=(Y_{1i},Y_{2i})$,  with ${\bm{Y}}_{i}={\bm{Y}}({\bm{X}}_i)$ and  $(Y_{1i},Y_{2i})=(Y_{1i}({\bm{X}}_i),Y_{2i}({\bm{X}}_i))$, can be decomposed as:
\begin{equation}
{\bm{Y}}_i=\bm{\mu}+{\bm{w}}_i, \quad i=1,\dots,n,\label{processdecomposedp}\end{equation}  
where  $\bm{\mu}=\bm{\mu}({\bm{X}}_i)\in\mathbb{R}^2$ is the mean vector and  ${\bm{w}}$ is a zero mean bivariate  Gaussian spatial process with cross covariance function $\rho_n({\bm{x}})\otimes {\bm{T}}$, being $\rho_n$ a continuous stationary correlation function and ${\bm{T}}$ a matrix, defined as:$${\bm{T}=\bigg(
 \begin{array}{cc}
 \sigma^2 & \tau\sigma\\
 \tau\sigma & 1
 \end{array}
 \bigg),}$$ with $\sigma>0$ and $\tau\in[-1,1]$. The operator $\otimes$ denotes the Kronecker product. A realization of a circular spatial error process (in vector form), $\bm{\varepsilon}=(\varepsilon_1,\dots,\varepsilon_n){^\top}$ is obtained as:
 $$\varepsilon_i={\mbox{arctan2}}(Y_{2i},Y_{1i}),\quad i=1,\dots,n.$$
 
 In the simulation study,   the projected Gaussian spatial process is generated setting $\bm{\mu}=(1,1){^\top}$ in (\ref{processdecomposedp}) (to ensure unimodality of the errors and thus obtain {homogeneous samples}) and considering the cross-covariance function:  \begin{equation}
 {\mathbb{C}\rm ov}({{\bm{w}}}_i,{\bm{w}}_j\mid{\bm{X}}_i, {\bm{X}}_j)=   
[\exp(-\norm{{\bm{X}}_i-{\bm{X}}_j}/a_\textrm{e})] {\bm{T}},
 \label{expo_p}
 \end{equation}where  $a_\textrm{e}$ is the practical
 range.  The variance and the parameter $\tau$ which controls the correlation between the linear variables are fixed to $\sigma=1$ and $\tau=0.9$, respectively, to better convey the dependence structure from the linear to the circular process \citep[see][Fig. 4]{wang2014modeling}. {As for the wrapping approach}, the realization of the circular error process $\bm{\varepsilon}$ must be centered. Different degrees of spatial dependence are
 studied, considering values of   $a_\textrm{e}=0.1$ (weak correlation), $a_\textrm{e}=0.3$ (medium correlation) and $a_\textrm{e}=0.6$ (strong correlation). Fig. \ref{figure:sim_proj}  shows a sample {on a $15\times 15$ grid} of a simulated projected Gaussian spatial process for different values of $a_\textrm{e}$, with values of $\bm{\mu}=(1,1){^\top}$, $\sigma=1$ and $\tau=0.9$. From left to right, the range increases, i.e., there is a stronger spatial dependence structure,   and, consequently,  the corresponding circular process realization shows a {smoother} pattern.

For the regression function $r_1$, numerical results are summarized in Tables \ref{table:simus_proj_R1b} and \ref{table:simus_proj_R1c}.   As it was pointed out in Section \ref{sec:wra} for the case of circular errors generated from  wrapped Gaussian spatial processes, {the CASE corresponding to the CV bandwidth matrix is the largest in all the scenarios.}   Regarding the MCV criterion,  when the dependence structure is stronger, the value of $l$ must be larger. For example, considering a weak dependence structure $(a_\textrm{e}=0.1)$ the use of ${\bm{H}}_{{\rm MCV}_{2}}$ seems to show a slightly better performance. If the dependence structure is stronger, ${\bm{H}}_{{\rm MCV}_{3}}$ provides better results.  Tables \ref{table:simus_proj_R2b} and  \ref{table:simus_proj_R2c} show the results for the regression function $r_2$, from where similar conclusions to those described when using $r_1$ can be deduced.

 Numerical outputs are completed with some additional plots. Given that similar results were obtained in the previous simulations for $\hat{m}_{{\bm{H}}}({\bm{x}};0)$ and $\hat{m}_{{\bm{H}}}({\bm{x}};1)$, plots are only shown for $\hat{m}_{{\bm{H}}}({\bm{x}};0)$. As an illustration of the appropriate performance of the estimator $\hat{m}_{{\bm{H}}}({\bm{x}};0)$, Fig. \ref{C_figure:regression} shows the theoretical regression functions    $r_1$ and $r_2$  (left panels) and
 the corresponding average, over 500 replicates, of the fitted values using $\hat{m}_{{\bm{H}}}({\bm{x}};0)$, considering samples of size $n=400$ and   circular errors  generated from a wrapped Gaussian spatial process (center panels) and from a projected Gaussian spatial process (right panels). In this example, for both types of circular errors, an exponential covariance model is used with range parameter equal to 0.3. Estimates are computed employing the bandwidth matrix ${\bm{H}}_{{\rm MCV}_{3}}$. Notice that, for comparison purposes, the theoretical regression functions are plotted  in a $30\times 30$ regular grid on the covariate region (the same grid where the estimations are computed).   Plots in the top row present the results for the data generated using the regression function  $r_1$ and those in the bottom row using $r_2$. The estimation of the circular trend surfaces seems to be quite accurate, no matter the approach ({wrapped} or projected) used to generate the circular spatial errors.
 
%----------------------------------------------------------%	
\section{Real data illustration}\label{sec:example}
%----------------------------------------------------------%
{The performance of the proposed estimators is illustrated on the Adriatic Sea wave direction dataset presented in the Introduction. A brief description of this dataset, as well as the regression model considered, are provided in Section \ref{sec:rd_des}. As noted in the Introduction, this dataset (or part of it) has been mainly analyzed using parametric methods \citep{jona2012spatial,wang2014modeling,lagona2015hidden,mastrantonio2016wrapped}, while we use now a nonparametric approach. Details on the kernel and bandwidth matrix employed in the estimation procedure are  given in Section \ref{sec:rd_bw}. Taking into account  that the performance of the regression estimators $\hat{m}_{{\bm{H}}}({\bm{x}};0)$ and $\hat{m}_{{\bm{H}}}({\bm{x}};1)$ was similar in the simulation study, only results employing $\hat{m}_{{\bm{H}}}({\bm{x}};1)$ are shown {for} this application. These results are included in Section \ref{sec:rd_est}. Some ideas on outliers diagnostics are provided in Section \ref{sec:rd_out}.}

\subsection{{Wave direction dataset and regression model}}\label{sec:rd_des}	
{Wave directions were recorded in 1494 grid points on the Adriatic Sea area from a calm period transitioning to a storm period at different times. These data outputs were derived from a wave model implemented by Istituto Superiore per la  Protezione e la Ricerca Ambientale (ISPRA) and they are available in  the \texttt{R} package \texttt{CircSpaceTime} \citep{CircSpaceTime}. }

In this illustration, we only consider wave directions for a calm period, corresponding to measurements taken at 06:00 on April 2 at Adriatic Sea (Fig. \ref{figure:realdata} shows a random sample of 150 observations of this dataset). We assume the  linear-circular regression model given in (\ref{model}), where  for $i=1,\dots,1494,$  ${\bm{X}}_{i}=(X_{i1},X_{i2}),$    represent the different locations, with $X_{i1}$ the longitude and $X_{i2}$ the latitude, and $\Theta_i$ the corresponding wave direction at that location.

	\subsection{{Bandwidth matrix selection}} \label{sec:rd_bw}
	  
\begin{figure}[t]
	\centerline{\includegraphics[width=8.5cm]{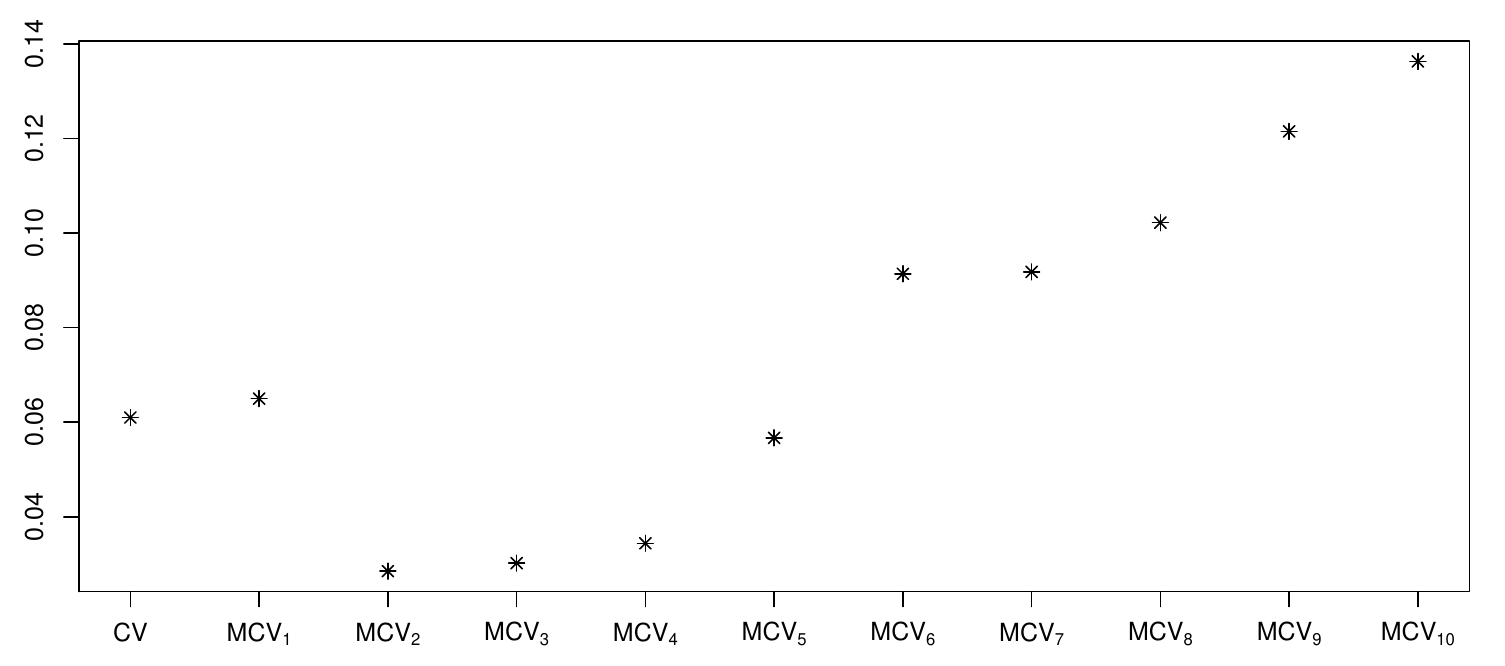}}
	\caption{\footnotesize Prediction errors given in (\ref{eq:error_rd}) for each bandwidth matrix selected by CV and {MCV, with different values of the radius $l(b)=\sqrt{2}b/10$, $b=1, \ldots, 10$.}}
	\label{figure:error_rd}
\end{figure}

	 {To compute the  nonparametric estimator of the circular regression function, given in (\ref{est}),  a multiplicative triweight kernel is considered.} The bandwidth matrix is selected employing a cross-validation criterion. In order to decide if using CV or {MCV} (and, in that case, a suitable value for {the radius} $l$), the whole sample is split in two parts, a randomly selected training sample of size $1345$ ($90\%$ of the data), denoted by $\{(\tilde{{\bm{X}}}_i,\tilde\Theta_i) \}_{i=1}^{1345}$, and a test sample, made up of the remaining observations, of size $149$ ($10\%$ of the data), denoted by $\{(\check{{\bm{X}}}_j,\check\Theta_j) \}_{j=1}^{149}$.  Then, estimations at each testing point $\check{{\bm{X}}}_j,j=1,\dots,149$, with different bandwidths, are compared with the testing responses using the following prediction error:
\begin{equation}\label{eq:error_rd}\sum_{j=1}^{149} \left\{1- \cos\left[\check{\Theta}_j - \hat{m}_{\hat{{\bm{H}}}}(\check{{\bm{X}}}_j;1)\right]\right\},
\end{equation}
where  $\hat{m}_{\hat{{\bm{H}}}}(\check{{\bm{X}}}_j;1)$ is the LL-type circular regression estimator  computed using the training sample and evaluated at the testing point $\check{{\bm{X}}}_j,j=1,\dots,149$, and $\hat{{\bm{H}}}$ denotes the bandwidth matrix selected using  CV or {MCV, employing the training sample. In the case of the MCV criterion,  different values of the radius $l$ are considered. As in the simulation study, we set $l(b)=\sqrt{2}b/10$, now with $b=1, \ldots, 10$.} These  bandwidth matrices are searched in the family of the symmetric and definite positive full bandwidth matrices, using an optimization algorithm based on the Nelder--Mead simplex method described in \cite{lagarias1998convergence}. To apply this optimization procedure, we use the  initial bandwidth matrix ${\bm{H}}_{{\text{init}}}=1.5 \cdot {\rm diag}\left\{\hat{\sigma}_{\tilde{X}_1} ,\hat{\sigma}_{\tilde{X}_2}  \right\}$, where  $\hat{\sigma}_{\tilde{X}_1}$ and $\hat{\sigma}_{\tilde{X}_2}$, with $\tilde{{\bm{X}}}=(\tilde{{X}}_{1},\tilde{{X}}_2)$,  are the training sample standard deviations of $\tilde{{X}}_{1}$ and $\tilde{{X}}_2$, respectively. Fig. \ref{figure:error_rd} shows the prediction error given in (\ref{eq:error_rd}) for each bandwidth matrix $\hat{{\bm{H}}}$. It can be seen that the minimum error is achieved when {MCV, with $b=2$,} is employed, converging the  algorithm  when using this criterion to
\begin{equation}\label{H_rd}
{\bm{H}}_{{\rm MCV}_{2}}=\left(\begin{array}{cc}
0.4744 & 0.0081 \\ 
0.0081 & 0.3529
\end{array} \right).
\end{equation}

{This solution is obtained from the application of a numerical optimization algorithm which, for our sample of relatively large size, took 4251.2983 seconds.}

\subsection{{Circular trend surface estimates}}\label{sec:rd_est}

\begin{figure} 
	\centerline{\includegraphics[width=9.1cm]{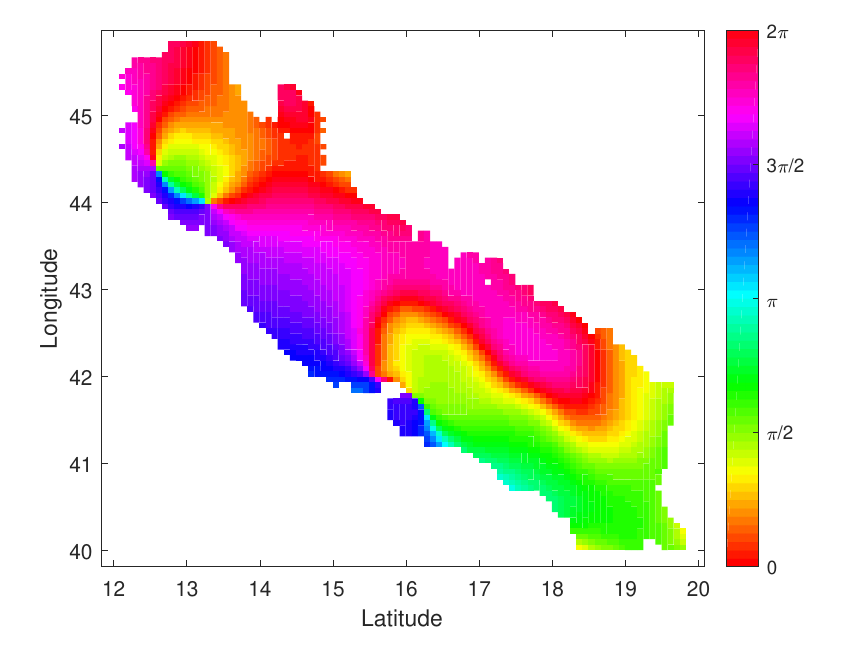}}
	\caption{\footnotesize Regression function estimation using the LL-type estimator $\hat{m}_{\hat{{\bm{H}}}}({\bm{x}};1)$, using the bandwidth matrix ${\bm{H}}$ given in (\ref{H_rd}), selected  with the   {MCV ($b=2$)} criterion.}
	\label{figure:rd}
\end{figure}

 The circular trend surface estimates using  ${\bm{H}}_{{\rm MCV}_{2}}$, given in (\ref{H_rd}), are shown in Fig. \ref{figure:rd}. The estimation grid is constructed by overlying the survey values of longitude and latitude with a $100 \times 100$ grid and, then, dropping every grid point that did
not satisfy at least one of the following two requirements: (a) it is within two ``grid cell length''  from an
observation point, or (b) the calculation for the estimates of the sine and cosine components at that grid point uses a smoothing vector that is sufficiently stable. {The sine and cosine of the detrended wave direction dataset were tested for isotropy and stationarity, following the proposals by \cite{bowman2013inference}. For both tests, $p$-values were larger than the usual significance levels (for isotropy: 0.3206 and 0.1271 for sine and cosine, respectively; for stationarity, $p$-values were larger than 0.99 for both processes)}.

From Fig.~\ref{figure:rd}, it can be clearly seen the shoreline orientation of the waves (recall that our measurements correspond to a calm period), providing the different {color pattern} along the coastline. Something which is interesting to notice is the behavior in the Gulf, where waves rotate to different directions, and a main current can be also observed. According to this pattern, more variation can be observed in the North, something that was also pointed out by \cite{jona2012spatial}, although for a storm period.

\subsection{Outlier {diagnostics}}\label{sec:rd_out}

\begin{figure}
\centering
\includegraphics[width=0.3\textwidth]{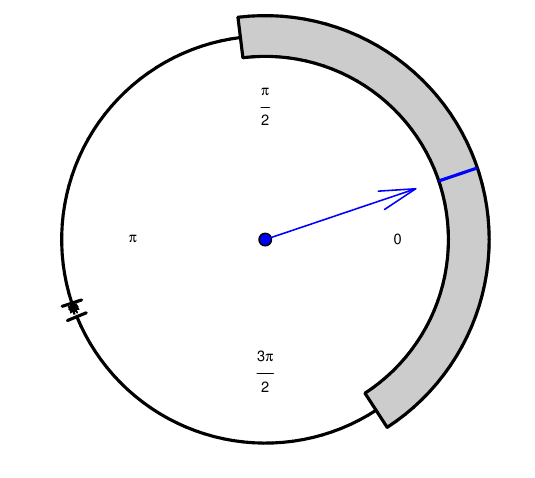}
\caption{ Circular boxplot of the residuals obtained from the proposed LL-type estimator.}
\label{outliers1}
\end{figure}

\begin{figure}
	\centering
	\includegraphics[width=0.25\textwidth]{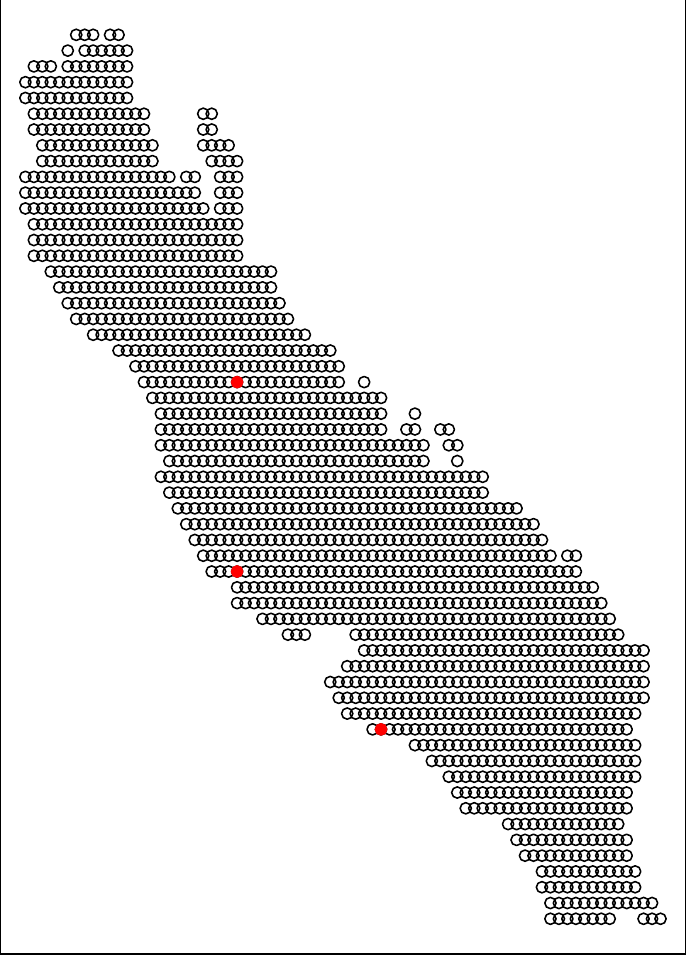}
	\caption{Spatial locations in the Adriatic Sea area on April 2, 2010 at 6am during a calm period (black points) and spatial locations which are (possible) circular outliers (red points).}
	\label{outliers2}
\end{figure}

{Diagnostics tools for outlier detection are required in order to round off the data modeling. In this context, a residual analysis seeking for possible outliers must be carried out using circular data tools. Apart from the initial ideas in \cite{jammalamadaka2001topics}, there have been some attempts to generalize Tukey's boxplot to the circular context, such as \cite{anderson1994graphical}, \cite{abuzaid2012boxplot} and more recently, \cite{buttarazzi2018boxplot} who devised a circular boxplot. Fig.~\ref{outliers1} shows the circular boxplot of the residuals from the nonparametric fit in Fig. \ref{figure:rd}, where just three outliers (corresponding to anomalous values of the residuals) are detected in our data sample. When plotting the locations corresponding to these values (Fig.~\ref{outliers2}), it can be observed that they are three isolated points. The LL-type estimator has been fitted again deleting these {three} points and, as expected, not much differences have been found. In this case, the minimum of the error, given in (\ref{eq:error_rd}), is achieved when {\rm MCV}  (with $b=2$) is employed, converging the  algorithm  when using this criterion to }

\begin{equation*}
{{\bm{H}}}_{{\rm MCV}_{2}}=\left(\begin{array}{cc}
0.5182 & 0.0065 \\ 
0.0065 & 0.4001\end{array} \right),
\end{equation*}
{yielding an almost identical estimation of the spatial trend to the one obtained with the whole dataset. This is actually expected given that this type of nonparametric estimators usually adapt satisfactorily in the presence of outliers. It should be noticed that if the circular boxplot is now recomputed when deleting the three points, no other outliers are identified.}

\section{{Conclusions, limitations, and further research}}
\label{sec:con} 
Nonparametric smoothing estimators of the regression function in a model with circular responses and real-valued covariates in the presence of spatial correlation are studied and applied to fit a circular trend surface for wave directions. The proposal considers  two  nonparametric regression models for the sine and cosine components of the circular responses, which are indeed regression models with real-valued res\-ponses. In particular,  NW and LL estimators are used in these two real-valued regression models. The asymptotic conditional properties of the proposed kernel-type  estimators are derived.
 
{Although there is a substantial literature on modeling circular data with spatial dependence by introducing and formulating spatial processes for circular data, such as wrapped and projected Gaussian spatial processes, 
our proposal follows a different perspective. We consider an appropriate linear-circular regression model for spatially correlated data and estimate nonparametrically the corresponding circular spatial trend. This is an alternative to model circular data at different spatial locations. It should be noted that no other direct competitor (up to our knowledge) has yet been proposed following these ideas, neither from a parametric nor from a nonparametric approach.}
	
{One of the advantages of the proposed procedure (if the bandwidth matrix is appropriately chosen, and a suitable bandwidth selector is also provided in this work) is that it relaxes parametric assumptions,  and consequently enables one to explore and model the data more 
flexibly, avoiding misspecification problems. Moreover, this estimator  can be employed as a first attempt to explore if a certain parametric family is appropriate or not to model the data.
It is well known that goodness-of-fit evaluation of a parametric regression mo\-del is often performed	using testing procedures, where the (parametric) fit of the model is compared to that obtained when considering a more general class of models. Nonparametric regression models are frequently employed in this setting \citep[for instance, ][]{hardle1993comparing,meilan2019goodness}.
Following these ideas, goodness-of-fit tests for assessing a parametric model for a regression model with circular response and $d$-dimensional covariate, in both independent and spatially correlated frameworks have been proposed and analyzed \citep{gof_circular}.}

{Regarding} the bandwidth matrix needed to compute the circular regression estimators given in (\ref{est}), it can be selected by leave-one-out cross-validation, but  this matrix is not necessarily a good one for spatially correlated data,  given that  ${\mathbb{E}}[{\rm CV}({\bm{H}})]$ is severely affected by the correlation   \citep{opsomer2001nonparametric,liu2001kernel}. In that context, it is advisable to employ other bandwidth selection criteria which take the spatial dependence structure into account. In our practical results, we also considered a modified cross-validation method suitably adjusted for
the presence of spatial correlation, which considers to ``leave $N(i)$ observations out''. The idea of modifying the selection criterion in this manner is not new. An example of such adjustment is the ``leave-$2l+1$-out'' cross-validation approach \citep{hart1990data}.  In the case of the marine currents in the Adriatic Sea,  a suitable parameter $l$ in {MCV, controlling} the number of observations left out, has been selected by minimizing the prediction error given in (\ref{eq:error_rd}).
{However, an interesting point would be to design a fully automatic procedure to compute the optimal radius in the MCV method. This approach should account for the spatial correlation of the covariates and, although it is out of the scope of the present paper, it would be an interesting topic of further research.
On the other hand,  cross-validation techniques have the drawback of being unable to provide satisfactory results in a reasonable time for very large sample sizes due to its computational complexity. To overcome this problem, bagging cross-validation bandwidths studied for density and regression estimation with Euclidean data \citep{HR} could be adapted to this context.
Additionally, note that even though cross-validation bandwidths present appealing theoretical properties, in practice, their computation could present certain difficulties in a multidimensional framework.}

Alternatively to the cross-validation methods previously described, the bandwidth could be selected as follows. First, as described before, the whole sample is split in two parts, a training and a testing samples. Then, using the overall dataset and a pilot bandwidth matrix, the nonparametric  estimator is computed at each training data point. From this estimation, and  using a wrapped \citep{jona2012spatial} or a projected Gaussian spatial process \citep{wang2014modeling}, the residuals can be modeled, obtaining predictions at the testing locations. Finally, the bandwidth matrix can be selected by minimizing the corresponding prediction error computed with the testing sample. 

{It should be noted that just global bandwidths are considered in the numerical studies of this paper. A limitation derived from the use of global bandwidths is that the corresponding nonparametric estimators may provide spurious estimates in areas with sparse observations. In such regions, the number of observations within the neighborhood determined by the bandwidth may be too small, producing unstable estimates. A way to overcome this problem is employing local bandwidths that automatically adapt to the number of observations near the grid points where calculating the estimates. The problem of using local bandwidths is that a high computing time would be required, specially if the sample size is very large.}

{Even though it seems we do not need to worry about outliers in our real data analysis, in a general case, deriving such a conclusion may be not as clear as for our example. Since the data exhibit spatial dependence, it is difficult to assess whether a set of observations are really outliers or just correspond to an effect of the spatial dependence structure.   Note that the definition of spatial outlier is not entirely precise and, besides, the {available} exploratory techniques {for real-valued spatial processes}, such as  the Moran scatterplot \citep{anselin1995local} or the variogram cloud \citep{cressie1993statistics}, do not allow a correct identification of such data. All this makes particularly difficult to {generalize the definition and the existing methods to} detect spatial outliers for circular data. There is certainly an interesting issue that could be addressed as further research.}

In the current setting, real-valued covariates for explaining the behavior of a circular response in presence of spatial correlation are considered. However,  it might be the case that other types of covariates, such  as other circular, or more generally, spherical covariates, may influence the circular response. For these more complex scenarios, there is a substantial research on modeling and on analyzing different inference approaches for random fields on spheres as well as on spheres across time. For instance,   \cite{porcu2016spatio}  developed cross-covariance functions of the great circle distances on the sphere. \cite{alegria2019covariance} proposed  a flexible parametric family of matrix-valued covariance functions. To overcome the problem of generating samples from random fields, \cite{emery2019simulating} introduced an algorithm to  generate isotropic vector-valued Gaussian random fields defined over the unit two-dimensional sphere embedded in the three-dimensional Euclidean space. Some of these approaches could be incorporated in our model, but these extensions are out of scope of the present paper and can be the focus of future researches.

 Finally, it is worth noting that environmental processes usually present an asymmetric behavior, requiring sophisticated distribution models (such as the Birn\-baum-Saunders (BS) distribution) for an appropriate fitting. In this context, \cite{saulo2013nonparametric} proposed a kernel method for estimating asymmetric densities based on a generalization of the BS model. From a regression perspective, \cite{leiva2020global} proposed a geostatistical model based on BS quantile regression and \cite{martinez2019birnbaum} formulated a regression model, considering a scalar response and functional covariates, supposing that its errors are spatially correlated and follow a BS distribution. In our context, no shape conditions are imposed regarding the (circular) error distribution, although the consideration of an extension of the BS distribution model to the circular context may enable the design of a parametric but relatively flexible regression model.

{In practice, the numerical studies performed in this work were run in an Intel Core i7-9700K at 3.60Ghz.  The simulations were implemented  in the statistical environment \texttt{R} \citep{Rsoft}, using functions included in the   \texttt{npsp} and \texttt{CircSpaceTime} packages  \citep{npsp, CircSpaceTime}. The real data application was performed in MATLAB software (www.mathworks.com).}

 \section*{Acknowledgements}\label{acknowledgements}
The authors acknowledge the support from the Xunta de Galicia grant ED481A-2017/361 and the European Union (European Social Fund - ESF). This research has been partially supported by MINECO grants  MTM2016-76969-P and MTM2017-82724-R, and by the Xunta de Galicia (Grupo de Referencia Competitiva  ED431C-2017-38, and Centro de Investigaci\'on del SUG ED431G 2019/01), all of them through the ERDF. The authors thank Prof. Agnese Panzera, from the University of Florence, for her help in the theoretical developments of the paper and her general comments about this work.
%{The authors also thank an Associate Editor and two anonymous referees for numerous useful comments that significantly improved this article.}

\bibliographystyle{spbasic}

\begin{thebibliography}{38}
	\providecommand{\natexlab}[1]{#1}
	\providecommand{\url}[1]{{#1}}
	\providecommand{\urlprefix}{URL }
	\expandafter\ifx\csname urlstyle\endcsname\relax
	\providecommand{\doi}[1]{DOI~\discretionary{}{}{}#1}\else
	\providecommand{\doi}{DOI~\discretionary{}{}{}\begingroup
		\urlstyle{rm}\Url}\fi
	\providecommand{\eprint}[2][]{\url{#2}}
	
	\bibitem[{Abuzaid et~al.(2012)Abuzaid, Mohamed, and
		Hussin}]{abuzaid2012boxplot}
	Abuzaid AH, Mohamed IB, Hussin AG (2012) Boxplot for circular variables.
	Computation Stat 27(3):381--392
	
	\bibitem[{Alegr{\'\i}a et~al.(2019)Alegr{\'\i}a, Porcu, Furrer, and
		Mateu}]{alegria2019covariance}
	Alegr{\'\i}a A, Porcu E, Furrer R, Mateu J (2019) Covariance functions for
	multivariate {G}aussian fields evolving temporally over planet earth. Stoch
	Env Res Risk A 33(8-9):1593--1608
	
	\bibitem[{Anderson(1994)}]{anderson1994graphical}
	Anderson CM (1994) Graphical methods for circular and cylindrical data. In:
	Tech. Report TR-94-05, Department of Statistical and Actuarial Sciences,
	University of Western Ontario
	
	\bibitem[{Anselin(1995)}]{anselin1995local}
	Anselin L (1995) Local indicators of spatial association-{LISA}. Geogr Anal
	27(2):93--115
	
	\bibitem[{Bowman and Crujeiras(2013)}]{bowman2013inference}
	Bowman AW, Crujeiras RM (2013) Inference for variograms. Comput Stat Data Anal
	66:19--31
	
	\bibitem[{Buttarazzi et~al.(2018)Buttarazzi, Pandolfo, and
		Porzio}]{buttarazzi2018boxplot}
	Buttarazzi D, Pandolfo G, Porzio GC (2018) A boxplot for circular data.
	Biometrics 74(4):1492--1501
	
	\bibitem[{Carnicero et~al.(2013)Carnicero, Aus{\'\i}n, and
		Wiper}]{carnicero2013non}
	Carnicero JA, Aus{\'\i}n MC, Wiper MP (2013) Non-parametric copulas for
	circular--linear and circular--circular data: an application to wind
	directions. Stoch Env Res Risk A 27(8):1991--2002
	
	\bibitem[{Casson and Coles(1998)}]{casson1998extreme}
	Casson E, Coles S (1998) Extreme hurricane wind speeds: estimation,
	extrapolation and spatial smoothing. J Wind Eng Ind Aerod 74:131--140
	
	\bibitem[{Cressie(1993)}]{cressie1993statistics}
	Cressie NA (1993) Statistics for spatial data. Wiley, New York
	
	\bibitem[{Di~Marzio et~al.(2012)Di~Marzio, Panzera, and Taylor}]{di2012non}
	Di~Marzio M, Panzera A, Taylor CC (2012) Non-parametric smoothing and
	prediction for nonlinear circular time series. J Time Ser Anal 33(4):620--630
	
	\bibitem[{Di~Marzio et~al.(2013)Di~Marzio, Panzera, and Taylor}]{di2013non}
	Di~Marzio M, Panzera A, Taylor CC (2013) Non-parametric regression for circular
	responses. Scand J Stat 40(2):238--255
	
	\bibitem[{Emery and Porcu(2019)}]{emery2019simulating}
	Emery X, Porcu E (2019) Simulating isotropic vector-valued {G}aussian random
	fields on the sphere through finite harmonics approximations. Stoch Env Res
	Risk A 33(8-9):1659--1667
	
	\bibitem[{Fern\'andez-Casal(2019)}]{npsp}
	Fern\'andez-Casal R (2019) {\texttt npsp}: Nonparametric spatial
	(geo)statistics. \urlprefix\url{http://cran.r-project.org/package=npsp}, {R}
	package version 0.7-5
	
	\bibitem[{Garc{\'\i}a-Portugu{\'e}s et~al.(2014)Garc{\'\i}a-Portugu{\'e}s,
		Barros, Crujeiras, Gonz{\'a}lez-Manteiga, and Pereira}]{garcia2014test}
	Garc{\'\i}a-Portugu{\'e}s E, Barros AM, Crujeiras RM, Gonz{\'a}lez-Manteiga W,
	Pereira J (2014) A test for directional-linear independence, with
	applications to wildfire orientation and size. Stoch Env Res Risk A
	28(5):1261--1275
	
	\bibitem[{Hall and Robinson(2009)}]{HR}
	Hall P, Robinson AP (2009) Reducing variability of crossvalidation for
	smoothing parameter choice. Biometrika 96:175--186
	
	\bibitem[{H\"ardle and Mammen(1993)}]{hardle1993comparing}
	H\"ardle W, Mammen E (1993) Comparing nonparametric versus parametric
	regression fits. Ann Stat 21:1926--1947
	
	\bibitem[{H\"{a}rdle and M\"{u}ller(2012)}]{hardle_muller}
	H\"{a}rdle W, M\"{u}ller M (2012) Multivariate and semiparametric kernel
	regression, John Wiley \& Sons, Ltd, chap~12, pp 357--391
	
	\bibitem[{Hart and Vieu(1990)}]{hart1990data}
	Hart JD, Vieu P (1990) Data-driven bandwidth choice for density estimation
	based on dependent data. Ann Stat 18(2):873--890
	
	\bibitem[{Jammalamadaka and Sengupta(2001)}]{jammalamadaka2001topics}
	Jammalamadaka SR, Sengupta A (2001) Topics in circular statistics, vol~5. World
	Scientific
	
	\bibitem[{Jona-Lasinio et~al.(2012)Jona-Lasinio, Gelfand, and
		Jona-Lasinio}]{jona2012spatial}
	Jona-Lasinio G, Gelfand A, Jona-Lasinio M (2012) Spatial analysis of wave
	direction data using wrapped {G}aussian processes. Ann Appl Stat
	6(4):1478--1498
	
	\bibitem[{Jona-Lasinio et~al.(2019)Jona-Lasinio, Mastrantonio, and
		Santoro}]{CircSpaceTime}
	Jona-Lasinio G, Mastrantonio G, Santoro M (2019) {\texttt CircSpaceTime}:
	spatial and spatio-temporal bayesian model for circular data.
	\urlprefix\url{http://cran.r-project.org/package=CircSpaceTime}, {R} package
	version 0.9.0
	
	\bibitem[{Kim and SenGupta(2017)}]{kim2017multivariate}
	Kim S, SenGupta A (2017) Multivariate-multiple circular regression. J Stat
	Comput Sim 87(7):1277--1291
	
	\bibitem[{Lagarias et~al.(1998)Lagarias, Reeds, Wright, and
		Wright}]{lagarias1998convergence}
	Lagarias JC, Reeds JA, Wright MH, Wright PE (1998) Convergence properties of
	the {N}elder--{M}ead simplex method in low dimensions. SIAM J Optimz
	9(1):112--147
	
	\bibitem[{Lagona et~al.(2015)Lagona, Picone, Maruotti, and
		Cosoli}]{lagona2015hidden}
	Lagona F, Picone M, Maruotti A, Cosoli S (2015) A hidden {M}arkov approach to
	the analysis of space--time environmental data with linear and circular
	components. Stoch Env Res Risk A 29(2):397--409
	
	\bibitem[{Leiva et~al.(2020)Leiva, S{\'a}nchez, Galea, and
		Saulo}]{leiva2020global}
	Leiva V, S{\'a}nchez L, Galea M, Saulo H (2020) Global and local diagnostic
	analytics for a geostatistical model based on a new approach to quantile
	regression. Stoch Env Res Risk A pp 1--15
	
	\bibitem[{Liu(2001)}]{liu2001kernel}
	Liu XH (2001) Kernel smoothing for spatially correlated data. PhD thesis,
	Department of Statistics, Iowa State University
	
	\bibitem[{Mart{\'\i}nez et~al.(2019)Mart{\'\i}nez, Giraldo, and
		Leiva}]{martinez2019birnbaum}
	Mart{\'\i}nez S, Giraldo R, Leiva V (2019) {B}irnbaum--{S}aunders functional
	regression models for spatial data. Stoch Env Res Risk A 33(10):1765--1780
	
	\bibitem[{Mastrantonio et~al.(2016)Mastrantonio, Gelfand, and
		Lasinio}]{mastrantonio2016wrapped}
	Mastrantonio G, Gelfand AE, Lasinio GJ (2016) The wrapped skew {G}aussian
	process for analyzing spatio-temporal data. Stoch Env Res Risk A
	30(8):2231--2242
	
	\bibitem[{Mastrantonio et~al.(2018)Mastrantonio, Pollice, and
		Fedele}]{mastrantonio2018distributions}
	Mastrantonio G, Pollice A, Fedele F (2018) Distributions-oriented wind forecast
	verification by a hidden {M}arkov model for multivariate circular--linear
	data. Stoch Env Res Risk A 32(1):169--181
	
	\bibitem[{Meil\'an-Vila et~al.(2020{\natexlab{a}})Meil\'an-Vila,
		Francisco-Fern\'andez, and Crujeiras}]{gof_circular}
	Meil\'an-Vila A, Francisco-Fern\'andez M, Crujeiras R (2020{\natexlab{a}})
	Goodness-of-fit tests for parametric regression models with circular
	response. arXiv: 2008.13473
	
	\bibitem[{Meil\'an-Vila et~al.(2020{\natexlab{b}})Meil\'an-Vila,
		Francisco-Fern\'andez, Crujeiras, and Panzera}]{meilan2019nonparametric}
	Meil\'an-Vila A, Francisco-Fern\'andez M, Crujeiras R, Panzera A
	(2020{\natexlab{b}}) Nonparametric multiple regression estimation for
	circular response. TEST \doi{10.1007/s11749-020-00736-w}
	
	\bibitem[{Meil\'an-Vila et~al.(2020{\natexlab{c}})Meil\'an-Vila, Opsomer,
		Francisco-Fern{\'a}ndez, and Crujeiras}]{meilan2019goodness}
	Meil\'an-Vila A, Opsomer JD, Francisco-Fern{\'a}ndez M, Crujeiras RM
	(2020{\natexlab{c}}) A goodness-of-fit test for regression models with
	spatially correlated errors. TEST 29:728--749
	
	\bibitem[{Opsomer et~al.(2001)Opsomer, Wang, and
		Yang}]{opsomer2001nonparametric}
	Opsomer J, Wang Y, Yang Y (2001) Nonparametric regression with correlated
	errors. Stat Sci 16:134--153
	
	\bibitem[{Porcu et~al.(2016)Porcu, Bevilacqua, and Genton}]{porcu2016spatio}
	Porcu E, Bevilacqua M, Genton MG (2016) Spatio-temporal covariance and
	cross-covariance functions of the great circle distance on a sphere. J Am
	Stat Assoc 111(514):888--898
	
	\bibitem[{{R Development Core Team}(2020)}]{Rsoft}
	{R Development Core Team} (2020) R: a language and environment for statistical
	computing. R Foundation for Statistical Computing, Vienna, Austria,
	\urlprefix\url{http://www.R-project.org}
	
	\bibitem[{Ruppert and Wand(1994)}]{ruppert1994multivariate}
	Ruppert D, Wand MP (1994) Multivariate locally weighted least squares
	regression. Ann Stat 22:1346--1370
	
	\bibitem[{Saulo et~al.(2013)Saulo, Leiva, Ziegelmann, and
		Marchant}]{saulo2013nonparametric}
	Saulo H, Leiva V, Ziegelmann FA, Marchant C (2013) A nonparametric method for
	estimating asymmetric densities based on skewed birnbaum--saunders
	distributions applied to environmental data. Stoch Env Res Risk A
	27(6):1479--1491
	
	\bibitem[{Wang and Gelfand(2014)}]{wang2014modeling}
	Wang F, Gelfand AE (2014) Modeling space and space-time directional data using
	projected {G}aussian processes. J Am Stat Assoc 109(508):1565--1580
	
\end{thebibliography}

\section*{Appendix. Proof of Theorem 1}
\begin{proof}
	
	Before deriving the asymptotic variance of  the estimator  $\hat{m}_{{\bm{H}}}({\bm{x}},p)$, for $p=0,1$, stated in Theorem 1, some preliminary approximations are needed. 
	
		Firstly, defining $f_1({\bm{x}})=\sin [m({\bm{x}})]$ and $f_2({\bm{x}})=\cos [m({\bm{x}})]$, using sine and cosine addition formulas, the following relation between the covariance function $C_{n,3}$, defined from models (\ref{model1}) and (\ref{model2}), and the correlations ${\rho_{k,n}}$, $k=1,2,3$, directly derived from model (\ref{model}) {and given in (\ref{cor1}), (\ref{cor2}) and (\ref{cor3})}, can be obtained:
			\begin{eqnarray*}
			{C}_{n,3}({\bm{X}}_i,{\bm{X}}_j)&=&f_1({\bm{X}}_i)f_2({\bm{X}}_j)\sigma^2_2{\rho_{2,n}}({\bm{X}}_i-{\bm{X}}_j)\\&&- f_1({\bm{X}}_i)f_1({\bm{X}}_j)\sigma_{12}{\rho_{3,n}}({\bm{X}}_i-{\bm{X}}_j)\\&&+ f_2({\bm{X}}_i)f_2({\bm{X}}_j)\sigma_{12}{\rho_{3,n}}({\bm{X}}_i-{\bm{X}}_j)\\&&- f_2({\bm{X}}_i)f_1({\bm{X}}_j)\sigma^2_1{\rho_{1,n}}({\bm{X}}_i-{\bm{X}}_j).
		\end{eqnarray*}

Moreover, denoting
	\begin{eqnarray*}k_{1,n}({\bm{x}})&=&		\dfrac{1}{n}\displaystyle\sum_{i=1}^{n} K_{{\bm{H}}}({\bm{X}}_i-{\bm{x}}),\\
	k_{2,n}({\bm{x}})&=& \dfrac{1}{n}\displaystyle\sum_{i=1}^{n} K_{{\bm{H}}}({\bm{X}}_i-{\bm{x}})({\bm{X}}_i-{\bm{x}}) ,\\
	k_{3,n}({\bm{x}})&=& \dfrac{1}{n}\displaystyle\sum_{i=1}^{n} K_{{\bm{H}}}({\bm{X}}_i-{\bm{x}})({\bm{X}}_i-{\bm{x}})({\bm{X}}_i-{\bm{x}}){^\top},\\	
	s_{1,n}({\bm{x}})&=&		\dfrac{1}{n^2}\bigg[\displaystyle\sum_{i=1}^n K^2_{{\bm{H}}}({\bm{X}}_i-{\bm{x}})c({\bm{X}}_i)+\displaystyle\sum_{i\neq j} K_{{\bm{H}}}({\bm{X}}_i-{\bm{x}})\\&&\cdot K_{{\bm{H}}}({\bm{X}}_j-{\bm{x}})	{C}_{n,3}({\bm{X}}_i,{\bm{X}}_j)\bigg],\\
	s_{2,n}({\bm{x}})&=&\dfrac{1}{n^2}\bigg[\displaystyle\sum_{i=1}^n K^2_{{\bm{H}}}({\bm{X}}_i-{\bm{x}})({\bm{X}}_i-{\bm{x}}) c({\bm{X}}_i)\\&&+\displaystyle\sum_{i\neq j} K_{{\bm{H}}}({\bm{X}}_i-{\bm{x}})K_{{\bm{H}}}({\bm{X}}_j-{\bm{x}})({\bm{X}}_i-{\bm{x}}) \\&&\cdot 	{C}_{n,3}({\bm{X}}_i,{\bm{X}}_j)\bigg],
\end{eqnarray*}	
	\begin{eqnarray*}
	s_{3,n}({\bm{x}})&=& \dfrac{1}{n^2}\bigg[\displaystyle\sum_{i=1}^n K^2_{{\bm{H}}}({\bm{X}}_i-{\bm{x}})({\bm{X}}_i-{\bm{x}})({\bm{X}}_j-{\bm{x}}) {^\top} c({\bm{X}}_i)\\&+&\displaystyle\sum_{i \neq j} K_{{\bm{H}}}({\bm{X}}_i-{\bm{x}})K_{{\bm{H}}}({\bm{X}}_j-{\bm{x}})({\bm{X}}_i-{\bm{x}})\\&&\cdot  ({\bm{X}}_j-{\bm{x}}) {^\top}	{C}_{n,3}({\bm{X}}_i,{\bm{X}}_j)\bigg],	
\end{eqnarray*}
and,	after some calculations, it can be obtained that
\begin{eqnarray}k_{1,n}({\bm{x}})&=&		f({\bm{x}})+o_{\mathbb{P}}(1) ,\label{k1n}\\
	k_{2,n}({\bm{x}})&=& {\mu_2}{\bm{\bm{\nabla}}} f({\bm{x}}){\bm{H}}^2+o_{\mathbb{P}}({\bm{H}}^2\bm{1}_d),\label{k2n}\\
	k_{3,n}({\bm{x}})&=&{\mu_2}f({\bm{x}}){\bm{H}}^2+o_{\mathbb{P}}({\bm{H}}\bm{1}_{d\times d}{\bm{H}}),\label{k3n}\\
%		s_{1,n}({\bm{x}})&=&\dfrac{1}{\abs{{\bm{H}}}}{\nu_0}f({\bm{x}})c({\bm{x}})+o_{\mathbb{P}}\left(\dfrac{1}{\abs{{\bm{H}}}}\right),\label{s1n}\\
	s_{1,n}({\bm{x}})&=&\dfrac{1}{n\abs{{\bm{H}}}}{\nu_0}f({\bm{x}})[c({\bm{x}})+f({\bm{x}})C_3({\bm{x}})]\nonumber\\&&+o_{\mathbb{P}}\left(\dfrac{1}{n\abs{{\bm{H}}}}\right),\label{s2n}\\
	s_{2,n}({\bm{x}})&=&\dfrac{1}{n\abs{{\bm{H}}}}o_{\mathbb{P}}({\bm{1}}_d)\label{s3n}\\
	s_{3,n}({\bm{x}})&=&		  \dfrac{1}{n\abs{{\bm{H}}}} o_{\mathbb{P}}({\bm{1}}_{d\times d}),	\label{s4n}
\end{eqnarray}
where $\bm{1}_d$ and $\bm{1}_{d\times d}$ denote the $d\times  1$ vector  and  the  $d\times  d$
matrix with every entry equal to 1, and 
	\begin{eqnarray}
	c({\bm{x}})&=&f_1({\bm{x}})f_2({\bm{x}})\sigma^2_2-f_1^2({\bm{x}})\sigma_{12}+f_2^2({\bm{x}})\sigma_{12}\nonumber\\&&-f_1({\bm{x}})f_2({\bm{x}})\sigma^2_1,\label{c}\\
	C_3({\bm{x}})&=&f_1({\bm{x}})f_2({\bm{x}})\sigma^2_2\rho_{{\textrm{c}}_2}-f_1^2({\bm{x}})\sigma_{12}\rho_{{\textrm{c}}_3}+ f_2^2({\bm{x}})\sigma_{12}\rho_{{\textrm{c}}_3}\nonumber\\&&-f_1({\bm{x}})f_2({\bm{x}})\sigma^2_1\rho_{{\textrm{c}}_1}.\label{c3}
\end{eqnarray}

To derive the variance of $\hat{m}_{{\bm{H}}}({\bm{x}};p)$, for $p=0,1$, denoting by $\bm{\mathcal{X}}=({\bm{X}}_1,\dots,{\bm{X}}_n)$, using Taylor expansions and following similar arguments to those used
in \cite{di2013non} and \cite{meilan2019nonparametric}, it can be obtained that
\begin{eqnarray}\lefteqn{{\mathbb{V}\rm ar}[\hat{m}_{{\bm{H}}}({\bm{x}};p)\mid\bm{\mathcal{X}}]\nonumber}\\&=&\dfrac{m_1^2({\bm{x}})}{\big[m_1^2({\bm{x}})+m_2^2({\bm{x}})\big]^2}{\mathbb{V}\rm ar}[\hat{m}_{2, {\bm{H}}}({\bm{x}};p)\mid\bm{\mathcal{X}}]\nonumber\\&&+ \dfrac{m_2^2({\bm{x}})}{\big[m_1^2({\bm{x}})+m_2^2({\bm{x}})\big]^2}{\mathbb{V}\rm ar}[\hat{m}_{1, {\bm{H}}}({\bm{x}};p)\mid\bm{\mathcal{X}}]\nonumber\\&&- \dfrac{2m_1({\bm{x}})m_2({\bm{x}})}{\big[m_1^2({\bm{x}})+m_2^2({\bm{x}})\big]^2}{\mathbb{C}\rm ov}[\hat{m}_{1, {\bm{H}}}({\bm{x}};p),\hat{m}_{2, {\bm{H}}}({\bm{x}};p)\mid\bm{\mathcal{X}}]\nonumber\\&&+ O\big\{[\hat{m}_{1, {\bm{H}}}({\bm{x}};p)-{m}_1({\bm{x}})]^3\big\}\nonumber\\&&+ O\big\{[\hat{m}_{2, {\bm{H}}}({\bm{x}};p)-{m}_2({\bm{x}})]^3\big\}.\label{var_est}\end{eqnarray}

	The conditional variance of $\hat{m}_{j, {\bm{H}}}({\bm{x}};p)$, for $j=1,2$, and $p=0,1,$ for spatially correlated data, can be derived using similar arguments to those given in \cite{liu2001kernel}, which yield
		\begin{eqnarray}\label{C_var} {\mathbb{V}\rm ar}[\hat m_{j, {\bm{H}}}({\bm{x}};p)\mid\bm{\mathcal{X}}]&=&\frac{{\nu_0}[s_j^2({\bm{x}})+f({\bm{x}})C_j({\bm{x}})]}{n \abs{{\bm{H}}}f({\bm{x}})}\nonumber\\&&+{o}_{\mathbb{P}}\left(\frac{1}{n \abs{{\bm{H}}}}\right),
		\end{eqnarray}
	where \begin{eqnarray}
		s_1^2({\bm{x}})&=&f_1^2({\bm{x}})\sigma^2_2+2f_1({\bm{x}})f_2({\bm{x}})\sigma_{12}+f_2^2({\bm{x}})\sigma^2_1,\label{s12}\\
	s_2^2({\bm{x}})&=&f_2^2({\bm{x}})\sigma^2_2-2f_2({\bm{x}})f_1({\bm{x}})\sigma_{12}+f_1^2({\bm{x}})\sigma^2_1,\label{s22}\\
		C_1({\bm{x}})&=&f_1^2({\bm{x}})\sigma^2_2\rho_{{\textrm{c}}_2}+2f_1({\bm{x}})f_2({\bm{x}})\sigma_{12}\rho_{{\textrm{c}}_3}\nonumber\\&&+ f_2^2({\bm{x}})\sigma^2_1\rho_{{\textrm{c}}_1},\label{c1}\\C_2({\bm{x}})&=&f_2^2({\bm{x}})\sigma^2_2\rho_{{\textrm{c}}_2}-2f_1({\bm{x}})f_2({\bm{x}})\sigma_{12}\rho_{{\textrm{c}}_3}\nonumber\\&&+ f_1^2({\bm{x}})\sigma^2_1\rho_{{\textrm{c}}_1}.\label{c2}
	\end{eqnarray}

	 Moreover, using (\ref{k1n}) and (\ref{s2n}), it is easy to obtain that the conditional covariance between $\hat{m}_{1, {\bm{H}}}({\bm{x}};0)$ and $\hat{m}_{2, {\bm{H}}}({\bm{x}};0)$ is
	\begin{eqnarray}\label{cov_m1_m2_v1}
	\lefteqn{{\mathbb{C}\rm ov}[\hat{m}_{1, {\bm{H}}}({\bm{x}};0),\hat{m}_{2, {\bm{H}}}({\bm{x}};0)\mid\bm{\mathcal{X}}]\nonumber}\\&=&\dfrac{\sum_{i=1}^{n}\sum_{j=1}^{n}K_{{\bm{H}}}({\bm{X}}_i-{\bm{x}})K_{{\bm{H}}}({\bm{X}}_j-{\bm{x}})}{\sum_{i=1}^{n}K_{{\bm{H}}}({\bm{X}}_i-{\bm{x}})\sum_{j=1}^{n}K_{{\bm{H}}}({\bm{X}}_j-{\bm{x}})}\nonumber\\&&\cdot {\mathbb{C}\rm ov}[\sin(\Theta_i),\cos(\Theta_j)\mid\bm{\mathcal{X}}]\nonumber\\&=& \dfrac{\sum_{i=1}^nK^2_{{\bm{H}}}({\bm{X}}_i-{\bm{x}})c({\bm{X}}_i)}{[\sum_{i=1}^{n}K_{{\bm{H}}}({\bm{X}}_i-{\bm{x}})]^2}\nonumber\\&&+ \dfrac{\sum_{i\neq j}K_{{\bm{H}}}({\bm{X}}_i-{\bm{x}})K_{{\bm{H}}}({\bm{X}}_j-{\bm{x}})C_{n,3}({\bm{X}}_i,{\bm{X}}_j)}{[\sum_{i=1}^{n}K_{{\bm{H}}}({\bm{X}}_i-{\bm{x}})]^2}\nonumber\\&=&\dfrac{1}{n\abs{{\bm{H}}}f({\bm{x}})}{\nu_0}[c({\bm{x}})+f({\bm{x}})C_3({\bm{x}})]\nonumber\\&&+ o_{\mathbb{P}}\left(\dfrac{1}{n\abs{{\bm{H}}}}\right).
\label{cov_m1_m2}
\end{eqnarray}

On the other hand,	the conditional covariance between $\hat{m}_{1, {\bm{H}}}({\bm{x}};1)$ and $\hat{m}_{2, {\bm{H}}}({\bm{x}};1)$ is
	\begin{eqnarray*}
		\lefteqn{{\mathbb{C}\rm ov}[\hat{m}_{1, {\bm{H}}}({\bm{x}};1),\hat{m}_{2, {\bm{H}}}({\bm{x}};1)\mid\bm{\mathcal{X}}]\nonumber}\\&=&{\bm{e}_1^\top(\bm{X}_{\bm{x}}^\top\bm{W}_{\bm{x}}\bm{X}_{\bm{x}})^{-1}\bm{X}_{\bm{x}}^\top\bm{W}_{\bm{x}}\bm{\Sigma} \bm{W}_{\bm{x}}\bm{X}_{\bm{x}}(\bm{X}_{\bm{x}}^\top\bm{W}_{\bm{x}}\bm{X}_{\bm{x}})^{-1}\bm{e}_1},
	\end{eqnarray*}
	where $\bm{\Sigma}$ is the covariance matrix of $\sin(\Theta)$ and $\cos(\Theta)$, whose $(i,j)$-entry is $\bm{\Sigma}(i,j)={\mathbb{C}\rm ov}[\sin(\Theta_i),\cos(\Theta_j)],$ $i,j=1,\dots,n.$ Using (\ref{k1n}), (\ref{k2n}), (\ref{k3n}), (\ref{s2n}), (\ref{s3n}) and (\ref{s4n}), it follows that 
	\begin{eqnarray*}
		\lefteqn{\left(n^{-1}{\bm{X}_{\bm{x}}^\top}{\bm{W}}_{{\bm{x}}}{\bm{X}}_{{\bm{x}}}\right)^{-1}}\\&=&\left(\begin{array}{ll}
			k_{1,n}({\bm{x}}) &  {k^\top_{2,n}(\bm{x})} \\
			k_{2,n}({\bm{x}}) & k_{3,n}({\bm{x}})
		\end{array}\right)^{-1}\\&=&
		\left(
		\begin{array}{ll}
			\frac{1}{f({\bm{x}})}+o_{\mathbb{P}}(1) & \frac{-\bm{\nabla}{^\top} f({\bm{x}})}{f^2({\bm{x}})}+o_{\mathbb{P}}({\bm{1}^\top_d}) \\
			\frac{-\bm{\nabla} f({\bm{x}})}{f^2({\bm{x}})}+o_{\mathbb{P}}({\bm{1}}_d)  & \frac{1}{{\mu_2}f({\bm{x}}){\bm{H}}^2}+o_{\mathbb{P}}({\bm{H}}{\bm{1}}_{d\times d}{\bm{H}})	\end{array}
		\right),\end{eqnarray*}	
	and that 
	\begin{eqnarray*}	\lefteqn{\dfrac{1}{n^2}{\bm{X}_{\bm{x}}^\top}{\bm{W}}_{{\bm{x}}}\bm{\Sigma}{\bm{W}}_{{\bm{x}}}	{\bm{X}}_{{\bm{x}}}}\\&=&\left(\begin{array}{ll}
			s_{1,n}({\bm{x}}) &  {s^\top_{2,n}(\bm{x})} \\
			s_{2,n}({\bm{x}}) & s_{3,n}({\bm{x}})
		\end{array}\right)\\&=&\dfrac{1}{n\abs{{\bm{H}}}}\left(
		\begin{array}{ll}
			{\nu_0}f({\bm{x}})[c({\bm{x}})+f({\bm{x}})C_3({\bm{x}})]{+o_{\mathbb{P}}(1)} & o_{\mathbb{P}}({\bm{1}^\top_d}) \\
			o_{\mathbb{P}}({\bm{1}}_d)  & o_{\mathbb{P}}({\bm{1}}_{d\times d})	\end{array}
		\right).\end{eqnarray*}

	Consequently, by straightforward calculations, one gets that
	\begin{eqnarray}
	\nonumber	&&{\mathbb{C}\rm ov}[\hat{m}_{1, {\bm{H}}}({\bm{x}};1),\hat{m}_{2, {\bm{H}}}({\bm{x}};1)\mid\bm{\mathcal{X}}]\\\nonumber&=&\dfrac{1}{n\abs{{\bm{H}}}f({\bm{x}})}{\nu_0}[c({\bm{x}})+f({\bm{x}})C_3({\bm{x}})]\\&&+ o_{\mathbb{P}}\left(\dfrac{1}{n\abs{{\bm{H}}}}\right).\label{cov_m1_m2_LL}
	\end{eqnarray}

 Using (\ref{var_est}), (\ref{C_var}), (\ref{cov_m1_m2}) and (\ref{cov_m1_m2_LL}),  one gets that, for $p=0,1$,
	\begin{eqnarray*}{\mathbb{V}\rm ar}[\hat{m}_{{\bm{H}}}({\bm{x}};p)\mid\bm{\mathcal{X}}]&=&\frac{1}{n \abs{{\bm{H}}}}\dfrac{{\nu_0}}{f({\bm{x}})}\dfrac{m_1^2({\bm{x}})}{[m_1^2({\bm{x}})+m_2^2({\bm{x}})]^2}\\&&\cdot [s_2^2({\bm{x}})+f({\bm{x}})C_2({\bm{x}})]\\&&+ \frac{1}{n \abs{{\bm{H}}}}\dfrac{{\nu_0}}{f({\bm{x}})}\dfrac{m_2^2({\bm{x}})}{[m_1^2({\bm{x}})+m_2^2({\bm{x}})]^2}\\&&\cdot [s_1^2({\bm{x}})+f({\bm{x}})C_1({\bm{x}})]\\&&- \dfrac{2}{n\abs{{\bm{H}}}}\dfrac{{\nu_0}}{f({\bm{x}})}\dfrac{m_1({\bm{x}})m_2({\bm{x}})}{[m_1^2({\bm{x}})+m_2^2({\bm{x}})]^2}\\&&\cdot [c({\bm{x}})+f({\bm{x}})C_3({\bm{x}})]+o_{\mathbb{P}}\bigg(\frac{1}{n \abs{{\bm{H}}}}\bigg).\end{eqnarray*}

Notice that	it holds that \begin{equation}\label{m1m2}
m_1({\bm{x}})=f_1({\bm{x}})\ell({\bm{x}})\quad\mbox{and}\quad
m_2({\bm{x}})=f_2({\bm{x}})\ell({\bm{x}}).
\end{equation}

	Taking into account that  $f^2_1({\bm{x}})+f_2^2({\bm{x}})=1$, it can be easily deduced that $\ell({\bm{x}})=[m^2_1({\bm{x}})+m_2^2({\bm{x}})]^{1/2}$. 	Therefore, using  (\ref{c}), (\ref{c3}), (\ref{s12}), (\ref{s22}), (\ref{c1}), (\ref{c2}) and (\ref{m1m2}),  it follows that
	\begin{eqnarray*}\label{l2sigma12}
	\lefteqn{m_1^2({\bm{x}})[s_2^2({\bm{x}})+f({\bm{x}})C_2({\bm{x}})]+m_2^2({\bm{x}})[s_1^2({\bm{x}})+f({\bm{x}})C_1({\bm{x}})]}\nonumber\\&&-2m_1({\bm{x}})m_2({\bm{x}})[c({\bm{x}})+f({\bm{x}})C_3({\bm{x}})]\nonumber\\&=&\ell^2({\bm{x}})\sigma^2_1[1+f({\bm{x}})\rho_{{\textrm{c}}_1}].
	\end{eqnarray*}
	
Consequently, it can be directly obtained that
		\begin{eqnarray*}{\mathbb{V}\rm ar}[\hat{m}_{{\bm{H}}}({\bm{x}};p)\mid\bm{\mathcal{X}}]&=&\dfrac{{\nu_0}\sigma^2_1[1+f({\bm{x}})\rho_{{\textrm{c}}_1}]}{n\abs{{\bm{H}}}\ell^2({\bm{x}})f({\bm{x}})}\\&&+ o_{\mathbb{P}}\bigg(\frac{1}{n \abs{{\bm{H}}}}\bigg).\end{eqnarray*}

\end{proof}

\end{document}